\documentstyle[12pt,epsf]{article}

\newcommand{\gsim}{\;\rlap{\lower 3.5 pt \hbox{$\mathchar \sim$}} \raise 1pt
 \hbox {$>$}\;}
\newcommand{\lsim}{\;\rlap{\lower 3.5 pt \hbox{$\mathchar \sim$}} \raise 1pt
 \hbox {$<$}\;}

\begin{document}
\noindent
\thispagestyle{empty}
\renewcommand{\thefootnote}{\fnsymbol{footnote}}
\begin{flushright}
{\bf TTP97-42}\footnote{The 
  complete postscript file of this
  preprint, including figures, is available via anonymous ftp at
  www-ttp.physik.uni-karlsruhe.de (129.13.102.139) as /ttp97-42/ttp97-42.ps 
  or via www at http://www-ttp.physik.uni-karlsruhe.de/cgi-bin/preprints.}
\hfill
{\bf UCSD/PHT 97-25}\\
{\bf MPI/PhT/97-65}
\hfill
{\bf DESY 97-220}\\
{\bf DTP/97/70}
\hfill
{\bf hep-ph/9711327}\\
{\bf November 1997}\\
\end{flushright}

\begin{center}
  \begin{Large}
Massive Quark Production in Electron Positron Annihilation to Order 
$\alpha_s^2$
\footnote{
Supported by BMBF under Contract 057KA92P, DFG under Contract Ku 502/8-1 
  and INTAS under Contract INTAS-93-744-ext.
}
  \end{Large}

  \vspace{0.5cm}

\begin{large}
 K.G.~Chetyrkin$^{a,b}$, 
 A.H.~Hoang$^{c}$,
 J.H.~K\"uhn$^{a}$,
 M.~Steinhauser$^{d}$
 and\\\vspace{2mm}
 T.~Teubner$^{e}$
\end{large}

\begin{center}
$^a$
   Institut f\"ur Theoretische Teilchenphysik, Universit\"at Karlsruhe,\\
   D-76128 Karlsruhe, Germany\\  
 \vspace{3mm}
$^b$
   Institute for Nuclear Research, Russian Academy of Sciences,\\
   Moscow 117312, Russia\\
 \vspace{3mm}
$^c$
   Department of Physics, University of California,\\
   San Diego, La Jolla, CA 92093-0319, USA\\ 
 \vspace{3mm}
$^d$
   Max-Planck-Institut f\"ur Physik, Werner-Heisenberg-Institut,\\
   D-80805 Munich, Germany\\
 \vspace{3mm}
$^e$
   Deutsches Elektronen-Synchrotron DESY,\\
   D-22603 Hamburg, Germany
\end{center}

  \vspace{0.7cm}
  {\bf Abstract}\\
\vspace{0.3cm}

\noindent
\begin{minipage}{15.0cm}
\begin{small}
Recent analytical and numerical results for the three-loop
polarization function allow to present a phenomenological analysis
of the cross section for massive quark production in electron positron
annihilation to order $\alpha_s^2$. Numerical predictions based on fixed
order perturbation theory are presented for charm and bottom
production above $5$ and $11.5$~GeV, respectively. The contribution
from these energy regions to $\alpha(M_Z^2)$, the running QED coupling
constant at scale $M_Z$, are given. The dominant terms close to threshold, 
i.e. in an expansion for small quark velocity $\beta$, are presented.
\end{small}
\end{minipage}

\end{center}

\setcounter{footnote}{0}
\renewcommand{\thefootnote}{\arabic{footnote}}
\vspace{1.2cm}

\thispagestyle{empty}
\newpage
\setcounter{page}{1}


\section{Introduction}
\label{sectionintro}

The total cross section for $e^+e^-$ annihilation into
hadrons, $\sigma_{had}$, constitutes one of
the most basic quantities of hadronic physics. It can be determined
experimentally and calculated theoretically with high precision.
It allows for a fundamental test of QCD and for a
precise determination of its parameters, the strong coupling constant, 
and the quark masses. In addition, it provides the decisive input for an
evaluation of the running QED coupling at high energies and for the
hadronic contribution to the lepton anomalous magnetic moment.
Perturbative QCD is expected to provide reliable predictions in the
continuum, i.e. one or two GeV above the respective quark threshold
and the respective resonance region.
Calculations in the massless limit have been performed 
several years ago in
${\cal O}(\alpha_s^2)$~\cite{Chet} and 
${\cal O}(\alpha_s^3)$~\cite{GorKatLar91SurSam91} which we also refer
to as NLO and NNLO
(for a review see~\cite{CKKRep}). The effect of nonvanishing quark 
masses, $M_Q$, has
been taken into account during the last years with the help of various
quite different approaches: a large number of
terms has been calculated in an expansion in 
$M_Q^2/s$~\cite{CheKue94,CheHarKueSte97},
where $\sqrt{s}$ is the center of mass energy,
and a subset of the
diagrams has been evaluated 
analytically~\cite{HJKT2,HKT1}. Alternatively, the
real and imaginary part of the polarization function $\Pi(q^2)$ 
has been obtained by
deriving analytical results for the expansions around
$q^2=0$, for the limit $M_Q^2/q^2\ll 1$ and around $q^2=4 M_Q^2$ and
by using the analyticity of $\Pi(q^2)$ to
reconstruct the full function numerically~\cite{CKS1}. 
The polarization function to order $\alpha_s^2$ for massive 
quarks is therefore under full control.

The previous papers were devoted to the technical aspects of the
calculation and to systematic tests and cross checks. The present paper
will be devoted to a compilation of the results in a simple and coherent
form and to various phenomenological applications. It contains, in
addition, the contribution from the double bubble diagram with a massive
quark in the internal and the external fermion loop. 
In the phenomenological applications the dominant
terms of order $\alpha_s^3$ in the massless 
approximation~\cite{GorKatLar91SurSam91} plus $M_Q^2/s$ terms~\cite{CheKue90}
will be included. This approach allows for a smooth
interpolation between the high energy region where the formulae are
accurate to NNLO order and the region closer to threshold  where
the results are valid to NLO order only.

The paper is organized as follows: a comprehensive
account of all NLO results is presented in Section~\ref{secpred}
for the sample case of the charm cross section.
In Section~\ref{seccbt} 
predictions for charm, bottom and top quark cross sections will
be given. The sensitivity of the results towards a variation of the
input parameters and the renormalization scale is investigated. In view
of their stability  in the regions away from the resonances the results
for the cross sections can be used to predict the contributions of the
charm and bottom continuum to the running of the QED coupling. A
detailed study of this effect is performed in Section~\ref{secqed}.  The
NLO perturbative results expanded for small velocities are 
an essential input for a
calculation of the cross section very close to threshold, i.e. for
energies comparable to or smaller than the Rydberg energy. In this region a
resummation of leading and subleading terms of order $\pi\alpha_s/\beta$,
where $\beta=\sqrt{1-4M_Q^2/s}$ is the velocity of the produced quarks,
is required. Although we do not perform this resummation of singular terms
in this paper, the essential ingredients from perturbation theory are
presented in Section~\ref{secthr}. 
Section~\ref{secsum} contains the summary
and conclusion.


\section{Predictions of order $\alpha_s^2$}
\label{secpred}

In a first step the theoretical results shall be recalled which are
required for the complete prediction of order $\alpha_s^2$
valid for energies sufficiently above the heavy quark threshold. The
crucial ingredient in the present approach is the existence of a
hierarchy of the quark masses. To be specific, let us consider the
region above the charm and below the bottom threshold --- the
generalization to the other cases of interest being obvious. The
energy is chosen sufficiently large, say above 5~GeV, to avoid the
complications in the regime very close to the $c\bar{c}$ threshold
(see Section~\ref{secthr}). The $u$, $d$ and $s$ quark masses are
neglected. Virtual bottom effects are treated through an expansion
in $s/(4M_b^2)$. This approximation is adequate in the full energy
region under consideration, even for 
$s/(4M_b^2) \to 1$~\cite{HJKT2}.
We are mainly
interested in the region where $\sqrt{s}$ and $M_c$ are of comparable
magnitude. It is thus convenient to identify the quark mass with the
pole mass, a convention adopted throughout this paper.

In order $\alpha_s^2$ the contributions from different quark species
to the vector current correlator
can be distinguished and thus added incoherently --- only ``non--singlet
terms'' are present in this order. 
The singlet contribution, which starts in order $\alpha_s^3$, 
has been calculated for massless quarks and is 
small~\cite{GorKatLar91SurSam91}.
In the following we shall first recall the
contributions arising from the electromagnetic current coupled to the
light $u$, $d$ and $s$ quarks, and subsequently the charm contribution
--- the main subject of this work.
We would like to stress, that the formulae, with the obvious replacements,
are equally well applicable for $b\bar{b}$ or $t\bar{t}$ production
above their respective thresholds (see Section~\ref{seccbt}).

\vspace{1em}

\noindent
1.~The sum of the absorptive parts of one-, two- and three-loop diagrams
with massless degrees of freedom (quarks or gluons) is given by~\cite{Chet}:
\begin{eqnarray}
R_{\rm light}(s) &=& 3 \sum_{i=u,d,s} Q_i^2\,\Bigg\{ 1 + 
\frac{\alpha_s^{(4)}(\mu^2)}{\pi} +
\left(\frac{\alpha_s^{(4)}(\mu^2)}{\pi}\right)^2 
\bigg[\frac{365}{24} - 11\,\zeta(3) \nonumber \\
&&\qquad\qquad\qquad \mbox{} +  n_{\ell} 
\left( -\frac{11}{12}+\frac{2}{3}\,\zeta(3)\right) + 
\left( -\frac{11}{4} + \frac{1}{6} n_{\ell} \right) 
 \ln\frac{s}{\mu^2} \bigg] \Bigg\}\,,
\label{eqrmassless}
\end{eqnarray}
where $\zeta(3)\approx1.2020569$ and
$n_{\ell} = n_f-1$ is the number of massless quarks. 
The $\overline{\rm MS}$ coupling $\alpha_s^{(4)}$
is to be evaluated at the scale $\mu^2$. Anticipating our strategy to
include the dominant $\alpha_s^3$ terms in the high energy region, the 
evolution of the strong coupling is governed by the three-loop beta 
function with $n_f = n_{\ell} + 1$ active flavours, where $n_f=4$
is chosen for the sample case of charm quark production. 

\begin{figure}[t]
\begin{center}
\begin{tabular}{ccc}
\leavevmode
\epsfxsize=5.0cm
\epsffile[142 267 470 525]{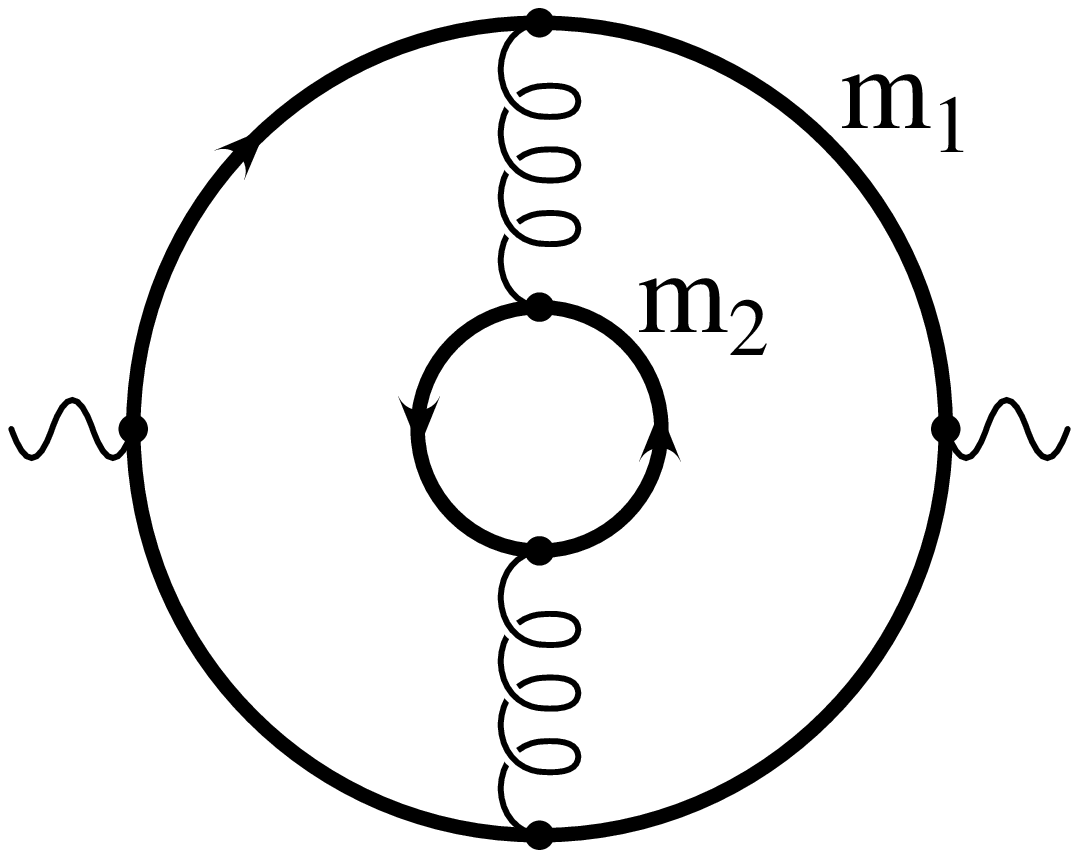}
&\hspace{2em}&
\leavevmode
\epsfxsize=5.0cm
\epsffile[142 267 470 525]{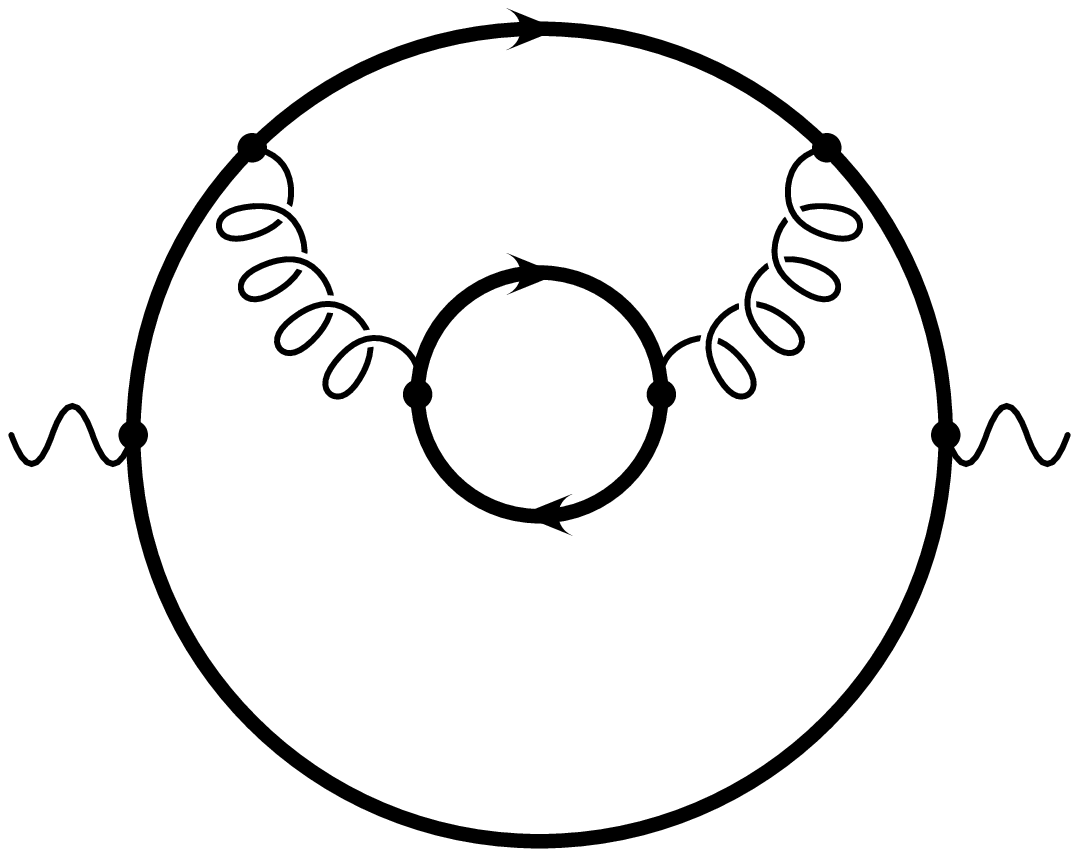}
\end{tabular}
\parbox{14.cm}{\small
\caption[]{\label{figgrcc} 
Fermionic double bubble diagrams with generic masses $m_1$ and $m_2$.
}}
\end{center}
\end{figure}

\vspace{1em}

\noindent
2.~Charm quarks can be produced through the splitting of gluons,
which in turn are radiated off $u$, $d$ or $s$ quarks. The analytic
result for this cross section can be found in~\cite{HJKT2}, the
corresponding virtual corrections to light quark pair production were
obtained earlier in~\cite{Kniehl} in the on-shell renormalization
scheme. The sum gives rise to the following ``double bubble''
contribution (see Fig.~\ref{figgrcc} with $m_1=0, m_2=M_c$):
\begin{eqnarray}
T C_F R_{qc}^{(2)} &=&
 3 \sum_{i=u,d,s} Q_i^2\,\frac{2}{3}\,\left(\rho^V(0,M_c^2,s) +
 \rho^R(0,M_c^2,s) + \frac{1}{4} \ln\frac{M_c^2}{\mu^2}\right)\,.
\label{eqr2qc}
\end{eqnarray}

\begin{figure}[ht]
\begin{center}
\begin{tabular}{c}
\leavevmode
\epsfxsize=13.cm
\epsffile[120 300 450 520]{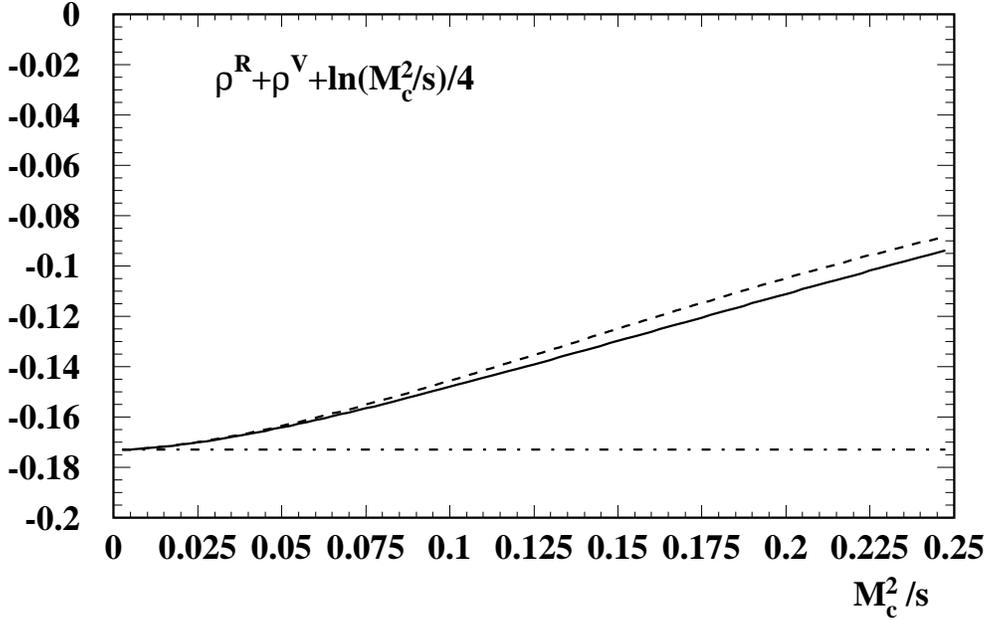}
\end{tabular}
\parbox{14.cm}{\small
\caption[]{\label{figqc} The function 
$\left(\rho^V(0,M_c^2,s) + \rho^R(0,M_c^2,s) 
+ \frac{1}{4}\ln\frac{M_c^2}{s}\right)$ as
described in the text. Solid line: exact result; dash-dotted line:
constant and quadratic terms only; 
dashed line: including terms up to order $M_c^4/s^2$ 
corrections (from~\cite{HJKT2}). The scale $\mu^2=s$ has been adopted.
}
}
\end{center}
\end{figure}

The functions $\rho^V(0,M_c^2,s)$ and $\rho^R(0,M_c^2,s)$ 
are given in~\cite{HJKT2}. The
combination on the right hand side of Eq.~(\ref{eqr2qc}) is well
approximated by the leading terms in the high energy limit. 
This is demonstrated in Fig.~\ref{figqc}, where the combination 
$\left(\rho^V(0,M_c^2,s) + \rho^R(0,M_c^2,s) 
+ \frac{1}{4}\ln\frac{M_c^2}{s}\right)$ is shown,
together with the leading terms of the high energy approximation 
($\mu^2=s$)
\begin{eqnarray}
\lefteqn{\rho^V(0,M_c^2,s) + \rho^R(0,M_c^2,s) 
+ \frac{1}{4}\ln\frac{M_c^2}{s} 
\,\,\,\stackrel{M_c^2/s\to0}{\longrightarrow} }
\nonumber\\&&
\zeta(3) - \frac{11}{8} 
+ \frac{M_c^4}{s^2}
  \left(-\frac{3}{2}\ln\frac{M_c^2}{s}- 6\,\zeta(3)+\frac{13}{2}\right)
+ {\cal O}\left(\frac{M_c^6}{s^3}\right).
\label{eqr2qchigh}
\end{eqnarray}
Note that the absence of $M_c^2$ terms could be inferred from general
renormalization group considerations~\cite{CheKue90}.

\vspace{1em}

\noindent
3.~Double bubble diagrams with external $u$, $d$, $s$ (or $c$) and
internal $b$ quarks decouple in the limit $s/(4M_b^2) \ll 1$. For
vanishing external quark mass an analytic result is 
available~\cite{Kniehl}, closely
related to $\rho^V(0,M_c^2,s)$ given above. It is well
approximated by the leading term in the $s/M_b^2$ 
expansion~\cite{Che93,SopSur94} ---
even up to $s = 4M_b^2$: 
\begin{equation}
TC_F \delta R_{qb}^{(2)} = 3 \sum_{i=u,d,s} Q_i^2
\,\frac{s}{4M_b^2}\,\bigg[\frac{8}{135}\ln\frac{M_b^2}{s}+\frac{176}{675}
\bigg]\,.
\label{eqrqb}
\end{equation}
These terms are numerically small.

\vspace{1em}

The combination of all terms proportional to $\sum_{i=u,d,s} Q_i^2$ 
thus reads
\begin{eqnarray}
R_{uds}(s) &=& 
R_{\rm light}
+\left(\frac{\alpha_s^{(4)}(\mu^2)}{\pi}\right)^2 T C_F
\left( R_{qc}^{(2)}(s) + \delta R_{qb}^{(2)}(s)\right).
\label{eqruds}
\end{eqnarray}
$R_{uds}$ is separately renormalization group invariant and approaches
$R_{\rm light}|_{n_\ell\to n_f}$ in the limit
$M_c^2\ll s\ll M_b^2$.

\vspace{1em}

Let us now proceed to the contributions arising from charm quarks
coupled to the electromagnetic current.
They will be cast into the form 
\begin{eqnarray}
R_c&=&Q_c^2\left(
R_c^{(0)} 
+ \frac{\alpha_s^{(4)}(\mu^2)}{\pi} \, C_F\,R_c^{(1)} 
+ \left(\frac{\alpha_s^{(4)}(\mu^2)}{\pi}\right)^2 \, R_c^{(2)}\,
\right).
\end{eqnarray}
The lowest order terms are well known~\cite{KaeSab55} and read
\begin{eqnarray}
R_c^{(0)}\,\,=\,\,3\,\beta\frac{3-\beta^2}{2}, 
&\qquad&
R_c^{(1)}\,\,=\,\,3\,\rho^{(1)},
\label{eqrc01}
\end{eqnarray}
where
\begin{eqnarray}
\rho^{(1)} &=&
 \frac{\left( 3 - {\beta^2} \right) \,\left( 1 + {\beta^2} \right) }{2
    }\,\bigg[\, 2\,\mbox{Li}_2(p) + \mbox{Li}_2({p^2}) + 
     \ln p\,\Big( 2\,\ln(1 - p) + \ln(1 + p) \Big) 
      \,\bigg]\,\nonumber\,\\ 
 & & \mbox{} - 
  \beta\,( 3 - {\beta^2} ) \,
   \Big( 2\,\ln(1 - p) + \ln(1 + p) \Big)  
\nonumber\\&&\mbox{}
  - 
  \frac{\left( 1 - \beta \right) \,
     \left( 33 - 39\,\beta - 17\,{\beta^2} + 7\,{\beta^3} \right) }{16}\,
   \ln p
+ 
  \frac{3\,\beta\,\left( 5 - 3\,{\beta^2} \right) }{8}\,,
\label{eqr1}
\end{eqnarray}
with
\begin{eqnarray}
  p \, = \, \frac{1-\beta}{1+\beta}\,,\qquad\,
    \beta \, = \, \sqrt{1-4M_c^2/s}\,
\end{eqnarray}
and $\mbox{Li}_n(p)$ is the polylogarithmic function.
In the limit $\beta\to0$ $\rho^{(1)}$ behaves as follows:
\begin{eqnarray}
\rho^{(1)}
&\stackrel{\beta\to0}{\longrightarrow}&
\frac{9}{2}\zeta(2)-6\beta+3\zeta(2)\beta^2+{\cal O}(\beta^3).
\end{eqnarray}
with $\zeta(2)=\pi^2/6$.
In order $\alpha_s^2$ a variety of diagrams has to be considered.

\vspace{1em}

\begin{figure}[t]
\begin{center}
\begin{tabular}{c}
\leavevmode
\epsfxsize=13.cm
\epsffile[110 270 470 540]{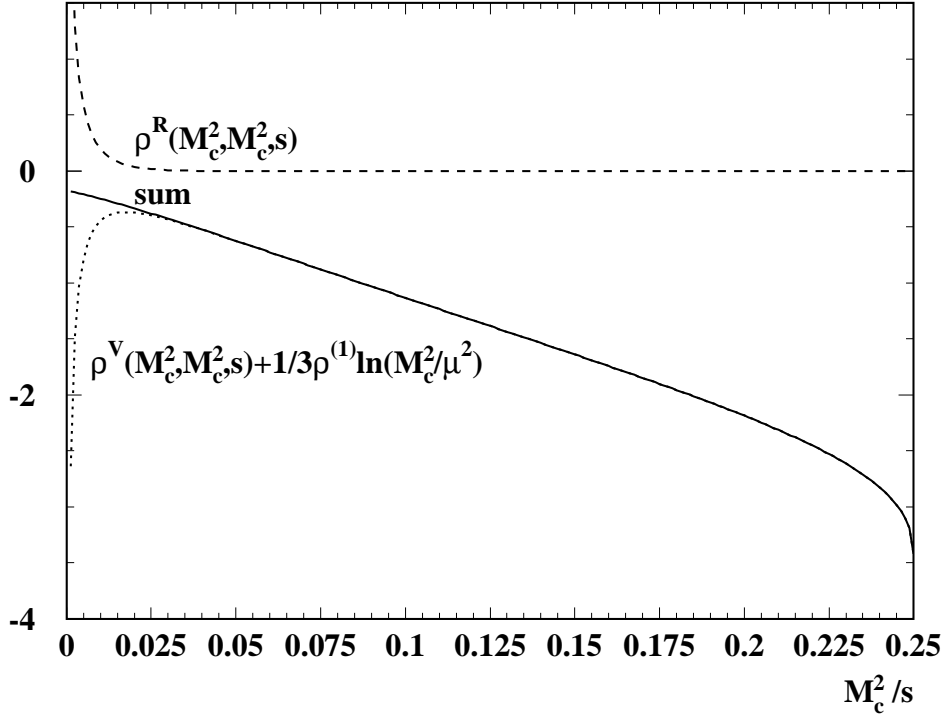}
\end{tabular}
\parbox{14.cm}{\small
\caption[]{\label{figcc} Second order contributions from double bubble
diagrams with massive internal and external quarks of the same mass
as a function of $M_c^2/s$. Dashed line: the real contribution 
$\rho^R(M_c^2,M_c^2,s)$; dotted line: the virtual correction 
$\rho^V(M_c^2,M_c^2,s) + \frac{1}{3}\rho^{(1)}\ln\frac{M_c^2}{\mu^2}$ 
for $\mu^2 = s$; solid line: the sum of both.
}}
\end{center}
\end{figure}

\noindent
4.~The essential ingredients for an evaluation of 
the double bubble diagram with two
charm quark loops (see Fig.~\ref{figgrcc} with $m_1=m_2=M_c$)
can be found in~\cite{HKT1}. 
The virtual corrections to
the $c\bar c$ vertex contribute for $\sqrt{s} > 2M_c$ and are known
analytically~\cite{HKT1}. The final state with four charm (anti--)
quarks is strongly suppressed close to its threshold at $4M_c \approx
7$~GeV. 
The rate is given in terms of a two dimensional integral to be
evaluated numerically. The combined contribution is thus written as 
\begin{eqnarray}
T\,C_F\,R_{cc}^{(2)} &=& 3\,\left(\frac{2}{3}\,
 \rho^V(M_c^2,M_c^2,s) + \frac{2}{3}\,\rho^R(M_c^2,M_c^2,s) + 
 \frac{1}{6} \ln\frac{M_c^2}{\mu^2} \,\frac{4}{3} \,\rho^{(1)} \right)
\label{eqr2cc}
\end{eqnarray}
where
\begin{eqnarray}
\rho^V(M_c^2,M_c^2,s) &=& \frac{1}{6} \,\Bigg[\, 
     \frac{3+10\beta^2-5\beta^4}{24} \,\ln^3 p +
     \frac{-3+40\beta^2+16\beta^4-15\beta^6}{12\beta^3} \,\ln^2 p  
 \nonumber \\
&&   \quad +\bigg( \frac{-18+234\beta^2+167\beta^4-118\beta^6}{18\beta^2} + 
            \frac{-3-10\beta^2+5\beta^4}{2} \,\zeta(2) \bigg) \ln p  
 \nonumber \\ 
&&   \quad +\frac{-9+510\beta^2-118\beta^4}{9\beta} + 
     \beta \,(-27+5\beta^2) \,\zeta(2) \, \Bigg]
\label{eqrccv}
\end{eqnarray}
and
\begin{equation}
\rho^R(M_c^2,M_c^2,s) = \frac{1}{3}\,
\int_{4M_c^2/s}^{(1-2M_c/\sqrt{s})^2}{\rm d}y\,
\int_{4M_c^2/s}^{(1-\sqrt{y})^2}\frac{{\rm d}z}{z}\,
\left(1+\frac{2M_c^2}{s z}\right)\sqrt{1-\frac{4M_c^2}{s z}}
\,\,{\cal F}(y,z)\,,
\label{eqrccr}
\end{equation}
with
\begin{eqnarray}
{\cal F}(y,z) & := &
\frac{\frac{8M_c^4}{s^2} + \frac{4M_c^2}{s}(1-y+z) - (1-y+z)^2 - 2(1+z)y}
 {1-y+z} \nonumber\\  
&&\cdot\ln\frac{1-y+z-\sqrt{1-\frac{4M_c^2}{s y}}\,\Lambda^{1/2}(1,y,z)}
        {1-y+z+\sqrt{1-\frac{4M_c^2}{s y}}\,\Lambda^{1/2}(1,y,z)}
\\[1mm]
&& -\sqrt{1-\frac{4M_c^2}{s y}}\,\Lambda^{1/2}(1,y,z)\, 
 \left[ 1 + \frac{ \frac{16M_c^4}{s^2}+\frac{8M_c^2}{s} + 
        4\left(1+\frac{2M_c^2}{s}\right)z }{
 (1-y+z)^2-\left(1-\frac{4M_c^2}{s y}\right)\,\Lambda(1,y,z)} \, \right] \,,
\nonumber\\[2mm]
\Lambda(1,y,z) &:=& 1+y^2+z^2-2(y+z+y z)\,.
\end{eqnarray}
Note that $\rho^R(M_c^2,M_c^2,s)$ vanishes for $s\to16M_c^2$.
The function $\rho^V(M_c^2,M_c^2,s)$ vanishes for $s\to4M_c^2$ and
$\rho^{(1)}$ approaches the constant $3\pi^2/4$ in the same limit. 
Both are zero below $4M_c^2$. For small $\beta$ one obtains:
\begin{eqnarray}
\rho^V(M_c^2,M_c^2,s)
&\stackrel{\beta\to0}{\longrightarrow}&
\left(\frac{22}{3}-4\zeta(2)\right)\beta 
+\left(-\frac{245}{54}+\frac{8}{3}\zeta(2)\right)\beta^3+{\cal O}(\beta^5)\,.
\end{eqnarray}
$R_{cc}^{(2)}$ is shown in Fig.~\ref{figcc} as a
function of $M_c^2/s$ in the range from 0 to 1/4. The contribution
from four particle production $\rho^R(M_c^2,M_c^2,s)$ and
the virtual correction 
$\rho^V(M_c^2,M_c^2,s)+\frac{1}{3}\rho^{(1)}\ln\frac{M_c^2}{s}$
are displayed separately as dashed and dotted lines, respectively,
their sum is shown as a solid line. 
In the high energy limit the sum approaches 
(for $\mu^2=s$) a constant value:
\begin{eqnarray}
\rho^V(M_c^2,M_c^2,s) + \rho^R(M_c^2,M_c^2,s) + 
\frac{1}{3}\rho^{(1)}\ln\frac{M_c^2}{s} 
&\stackrel{M_c^2/s\to0}{\longrightarrow}&
\zeta(3) - \frac{11}{8}
+{\cal O}\left(\frac{M_c^2}{s}\right)
\,.
\label{eqccasym}
\end{eqnarray}
Fig.~\ref{figcc} demonstrates that for energies far above the four
particle threshold, i.e. for $\sqrt{s} \gg 4M_c$, 
real and virtual contributions 
cancel to a large extent. For smaller energies, however, the
(negative) virtual corrections become increasingly more important
as the energy decreases down to the two particle threshold.

\vspace{1em}

The contributions from charm quarks coupled to the external current
with internal massless quark or gluon lines are significantly more
important. Their treatment is the main subject of this paper. Very close 
to threshold the Coulomb singularity has to be incorporated and the 
definition of the coupling has to be scrutinized. However, in a first step, 
the energy region will be considered where mass terms are important but 
Coulomb resummation is not yet required.

\vspace{1em}

\noindent
5.~Let us start with double bubble diagrams with light internal quark loops
(see Fig.~\ref{figgrcc} with $m_1=M_c, m_2=0$). In the previous cases,
Eqs.~(\ref{eqr2qc}) and (\ref{eqr2cc}) with massive internal quark
loops, the rates for real and virtual radiation could be given
separately and no mass singularity was present. This differs from the
case with vanishing internal quark mass: quadratic and linear mass 
logarithms arise in the individual cuts which can be cancelled by
combining real and virtual emission and by adopting the $\overline{\rm
MS}$ definition of the coupling constant. The analytical result has
been obtained in~\cite{HKT1} (see also~\cite{HoaTeu97}). 
For completeness we recall the result 
for $n_{\ell}$ light quark species:
\begin{eqnarray}
n_{\ell}\,T\,C_F\,R_{cq}^{(2)} &=& 3\,\frac{2}{3}
\,n_{\ell}\left(
-\,\frac{1}{3}\,\bigg[\ln\frac{\mu^2}{s} + \frac{5}{3} - \ln 4 \bigg]\,
\rho^{(1)} + \delta^{(2)}\right)\,,
\label{eqr2cq}
\end{eqnarray}
where
$\rho^{(1)}$ is given in Eq.~(\ref{eqr1}) and 
\begin{eqnarray}
\lefteqn{
\delta^{(2)} \, = \,
- \,\frac{\left( 3 - {{\beta }^2} \right) \,
       \left( 1 + {{\beta }^2} \right) }{6}\,}\nonumber\,\\ 
 & & \mbox{}\,
     \qquad\,\cdot\Bigg\{\,\mbox{Li}_3(p) - 2\,\mbox{Li}_3(1 - p) - 
       3\,\mbox{Li}_3({p^2}) - 4\,\mbox{Li}_3\Big({p\over {1 + p}}\Big) - 
       5\,\mbox{Li}_3(1 - {p^2}) + 
       \frac{11}{2}\,\zeta(3)\,\nonumber\,\\ 
 & & \mbox{}\,\qquad + 
       \mbox{Li}_2(p)\,\ln\Big(\frac{4\,\left( 1 - {{\beta }^2} \right) }{
          {{\beta }^4}}\Big) + 2\,\mbox{Li}_2({p^2})\,
        \ln\Big(\frac{1 - {{\beta }^2}}{2\,{{\beta }^2}}\Big) + 
       2\,\zeta(2)\,\bigg[\, \ln p - 
          \ln\Big(\frac{1 - {{\beta }^2}}{4\,\beta }\Big) \,\bigg] 
      \,\nonumber\,
        \\ 
 & & \mbox{}\,\qquad - 
       \frac{1}{6}\,\ln\Big(\frac{1 + \beta }{2}\Big)\,
        \bigg[\, 36\,\ln 2\,\ln p - 44\,\ln^2 p + 
          49\,\ln p\,\ln\Big(\frac{1 - {{\beta }^2}}{4}\Big) + 
          \ln^2\Big(\frac{1 - {{\beta }^2}}{4}\Big) \,\bigg] \,\nonumber\,
        \\ 
 & & \mbox{}\,\qquad - 
       \frac{1}{2}\,\ln p\,\ln \beta\,
        \bigg[\, 36\,\ln 2 + 21\,\ln p + 16\,\ln \beta  - 
          22\,\ln(1 - {{\beta }^2}) \,\bigg]  \,\Bigg\} \,\nonumber\,
     \\ 
 & & \mbox{}   + 
  \frac{1}{24}\,\Bigg\{ \,
      ( 15 - 6\,{{\beta }^2} - {{\beta }^4} ) \,
      \Big( \mbox{Li}_2(p) + \mbox{Li}_2({p^2}) \Big)  + 
     3\,( 7 - 22\,{{\beta }^2} + 7\,{{\beta }^4} ) \,
      \mbox{Li}_2(p)\,\nonumber\,\\ 
 & & \mbox{}\,\qquad - 
     ( 1 - \beta  ) \,
      ( 51 - 45\,\beta  - 27\,{{\beta }^2} + 5\,{{\beta }^3} ) \,
      \zeta(2)\,\nonumber\,\\[2mm] 
 & & \mbox{}\,\qquad + 
     \frac{\left( 1 + \beta  \right)}{\beta} \,
        \left( -9 + 33\,\beta  - 9\,{{\beta }^2} - 15\,{{\beta }^3} + 
          4\,{{\beta }^4} \right) \,\ln^2 p\,\nonumber\,
      \\ 
 & & \mbox{}\,\qquad + 
     \bigg[ \,( 33 + 22\,{{\beta }^2} - 7\,{{\beta }^4} ) \,
         \ln 2 - 10\,( 3 - {{\beta }^2} ) \,
         ( 1 + {{\beta }^2} ) \,\ln \beta \,
   \nonumber\,    \\ 
 & & \mbox{}\,\qquad\,\qquad\,\qquad   - 
        ( 15 - 22\,{{\beta }^2} + 3\,{{\beta }^4} ) \,
         \ln\Big(\frac{1 - {{\beta }^2}}{4\,{{\beta }^2}}\Big)\,\bigg] \,
      \ln p\,\nonumber\,\\ 
 & & \mbox{}\,\qquad + 
     2\,\beta \,( 3 - {{\beta }^2} ) \,
   \ln\Big(\frac{4\,\left( 1 - {{\beta }^2} \right) }{{{\beta }^4}}\Big)\,
   \bigg[\, \ln \beta - 3\,\ln\Big(\frac{1 - {{\beta }^2}}{4\,\beta }\Big) 
      \,\bigg]
       \,\nonumber\,\\ 
 & & \mbox{}\,\qquad + 
     \frac{237 - 96\,\beta  + 62\,{{\beta }^2} + 32\,{{\beta }^3} - 
        59\,{{\beta }^4}}{4}\,\ln p - 
     16\,\beta \,( 3 - {{\beta }^2} ) \,\ln\Big(\frac{1 + \beta }{4}\Big)\,
      \nonumber\,\\ 
 & & \mbox{}\,\qquad - 
     2\,\beta \,( 39 - 17\,{{\beta }^2} ) \,
      \ln\Big(\frac{1 - {{\beta }^2}}{2\,{{\beta }^2}}\Big) - 
     \frac{\beta \,\left( 75 - 29\,{{\beta }^2} \right) }{2}\,\Bigg\} 
\,.
\label{deltatwoloop}
\end{eqnarray}
For small velocities $\delta^{(2)}$ is given by:
\begin{eqnarray}
\delta^{(2)}
&\stackrel{\beta\to0}{\longrightarrow}&
  3\zeta(2)\,\ln \frac{\beta}{2}  + 
  \left( -\frac{3}{2} + 8\,\ln 2 \right) \,\beta + 
  2\zeta(2)\,\left( \ln \frac{\beta}{2} - 2 \right) \,{\beta^2}
 \,+\,{\cal{O}}(\beta^3)
\,.
\end{eqnarray}

\vspace{1em}

\begin{figure}[t]
\begin{center}
\begin{tabular}{ccc}
\leavevmode
\epsfxsize=5.0cm
\epsffile[142 267 470 525]{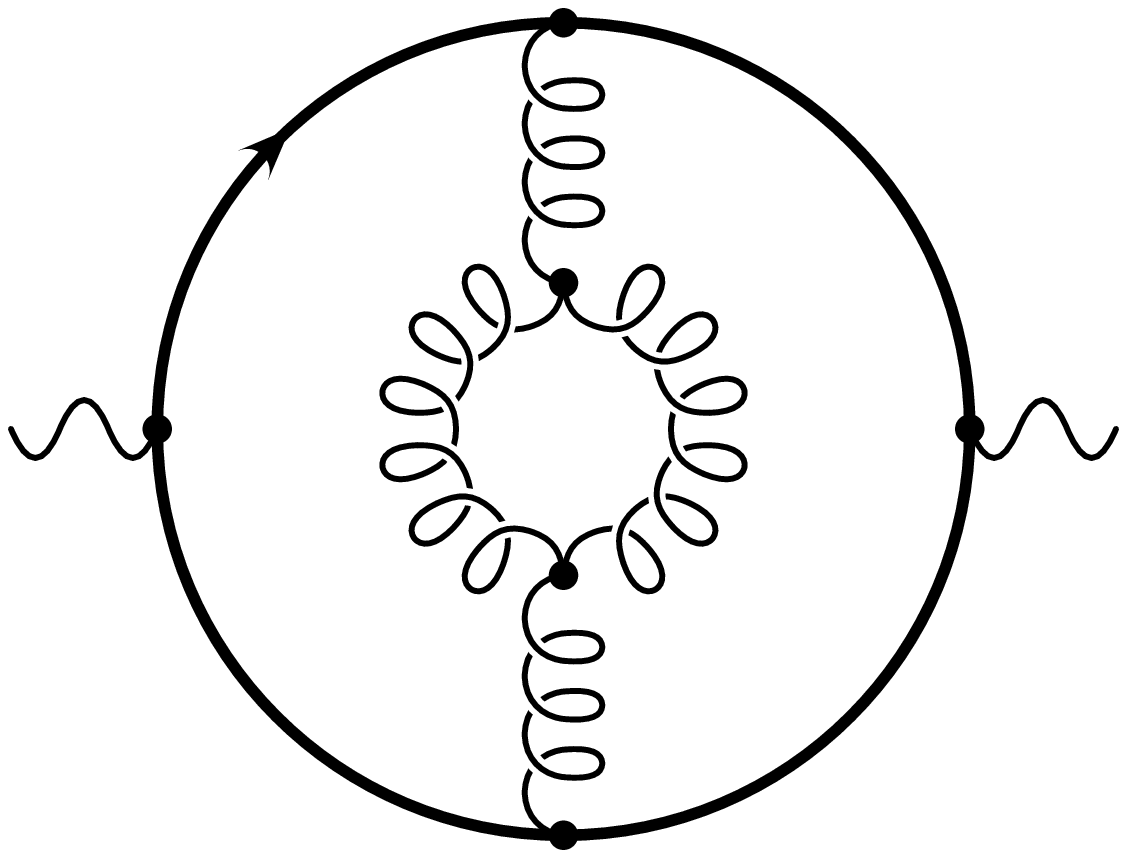}
&\hspace{2em}&
\leavevmode
\epsfxsize=5.0cm
\epsffile[142 267 470 525]{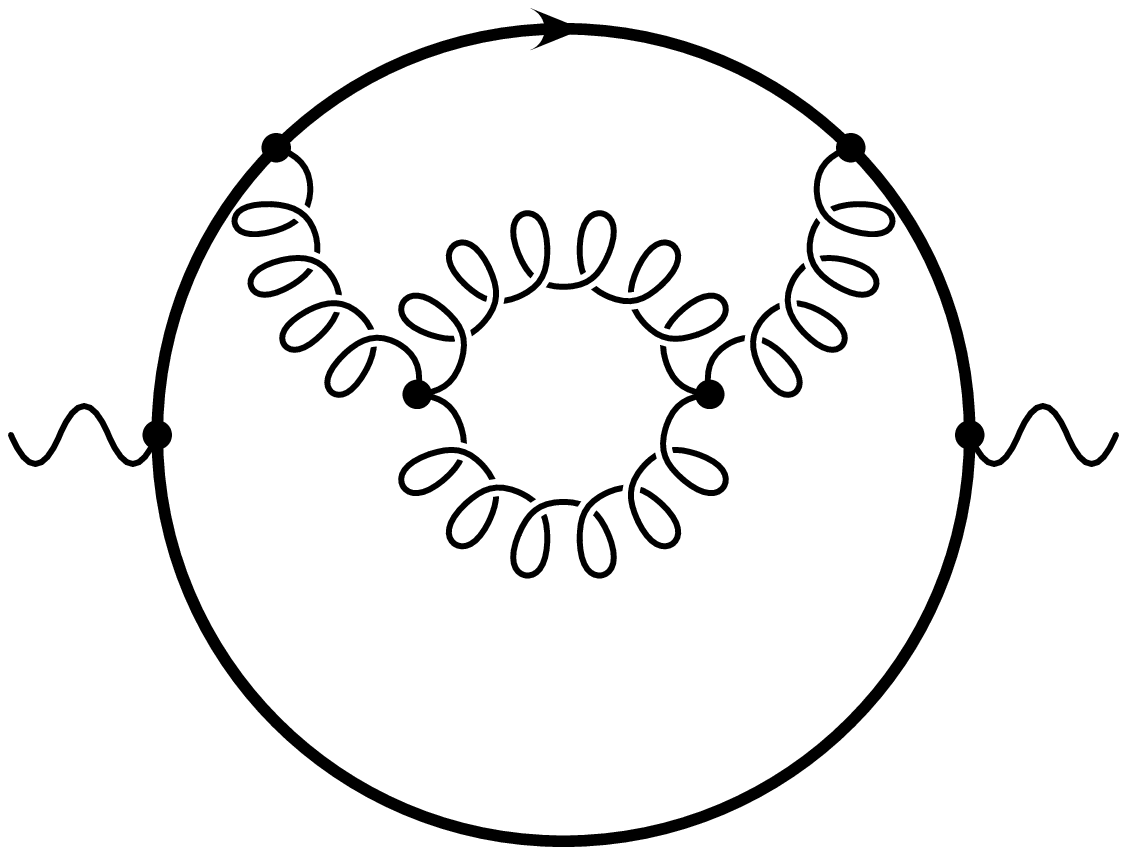}
\end{tabular}
\parbox{14.cm}{\small
\caption[]{\label{figgrcg} Gluonic double bubble diagrams where the 
external photon is coupled to a heavy quark and the emitted gluon 
splits into a gluon loop. The diagrams for the ghost particles
which also have to be taken into account, are not depicted.
}}
\end{center}
\end{figure}

\noindent
6.~Diagrams with massive quarks and purely gluonic internal lines have
been evaluated in~\cite{CKS1} through a combination of analytical and
numerical methods. The decomposition of the result according to the
colour structure will be important for the discussion in 
Section~\ref{secthr} 
below. Terms proportional to $C_F^2$ with a threshold singularity
proportional to 
$(\pi\alpha_s)^2/\beta$ are present in abelian and nonabelian
theories as well, whereas terms proportional to $C_F C_A$ with a logarithmic
threshold singularity are characteristic for the nonabelian structure 
of the theory, with a behaviour similar to the $C_F T n_{\ell}$ term. 
This leads to the decomposition 
\begin{equation}
R_c^{(2)} = C_F^2 R_A^{(2)} + C_F C_A R_{NA}^{(2)} +
T C_F n_{\ell} R_{c q}^{(2)} 
+ T C_F R_{c c}^{(2)}
+ T C_F \delta R_{cb}^{(2)}\,,
\label{eqdecomprtwo}
\end{equation}
where $\delta R_{cb}^{(2)}$ is the contribution with an internal 
$b$ quark loop obtained in analogy to Eq.~(\ref{eqrqb}).

The following approximations have been derived in~\cite{CKS1}:
\begin{eqnarray}
R_A^{(2)} &=& \frac{(1-\beta^2)^4}{\beta}\,\frac{3\pi^4}{8}
              - 12 \,\rho^{(1)}
                +\beta\,\frac{2619}{64}-\beta^3\,\frac{2061}{64}
                +\frac{81}{8}\left(1-\beta^2\right)\ln p
\nonumber\\
&& - 198\left(\frac{M_c^2}{s}\right)^{3/2} \left(\beta^4-2\beta^2\right)^6 
\nonumber\\
&& + 100 \,p^{3/2} (1-p) \left(2.08 - 1.57 \,p + 0.405 \,p^2\right)\,,
\label{appfora}
\\
R_{NA}^{(2)} &=& R_{g}^{(2)}\Big|_{\xi=4}
               + \beta\,\frac{351}{32} - \beta^3\,\frac{297}{32}
\nonumber\\
&& - 18\left(\frac{M_c^2}{s}\right)^{3/2}  \left(\beta^4-2\beta^2\right)^4
\nonumber\\
&& + 50 \,p^{3/2} (1-p) \left(1.41 - 1.24 \,p + 0.96 \,p^2\right)\,,
\label{appfornaxi}
\end{eqnarray}
where $p=(1-\beta)/(1+\beta)$. $R_{g}^{(2)}|_{\xi=4}$ is the 
contribution from gluonic double bubble diagrams (see Fig.~\ref{figgrcg}) 
for the special choice of the gauge parameter $\xi=4$ and reads~\cite{CHKST1}
\begin{equation}
R_{g}^{(2)}\Big|_{\xi=4} = \left( \frac{11}{4} \ln\frac{\mu^2}{4s} + 
 \frac{31}{12} \right) \rho^{(1)} - \frac{33}{4} \,\delta^{(2)}\,, 
\label{eqrg}
\end{equation}
with $\rho^{(1)}$ and $\delta^{(2)}$ given in Eqs.~(\ref{eqr1}) and 
(\ref{deltatwoloop}), respectively. The first lines of 
Eqs.~(\ref{appfora}) and (\ref{appfornaxi}) 
consist of the exactly known high energy and threshold
contributions, the second and third lines represent a numerically small
reminder.


\section{Cross section for the heavy quark production}
\label{seccbt}

\begin{figure}[b]
\begin{center}
\begin{tabular}{cc}
\leavevmode
\epsfxsize=7.0cm
\epsffile[100 280 460 560]{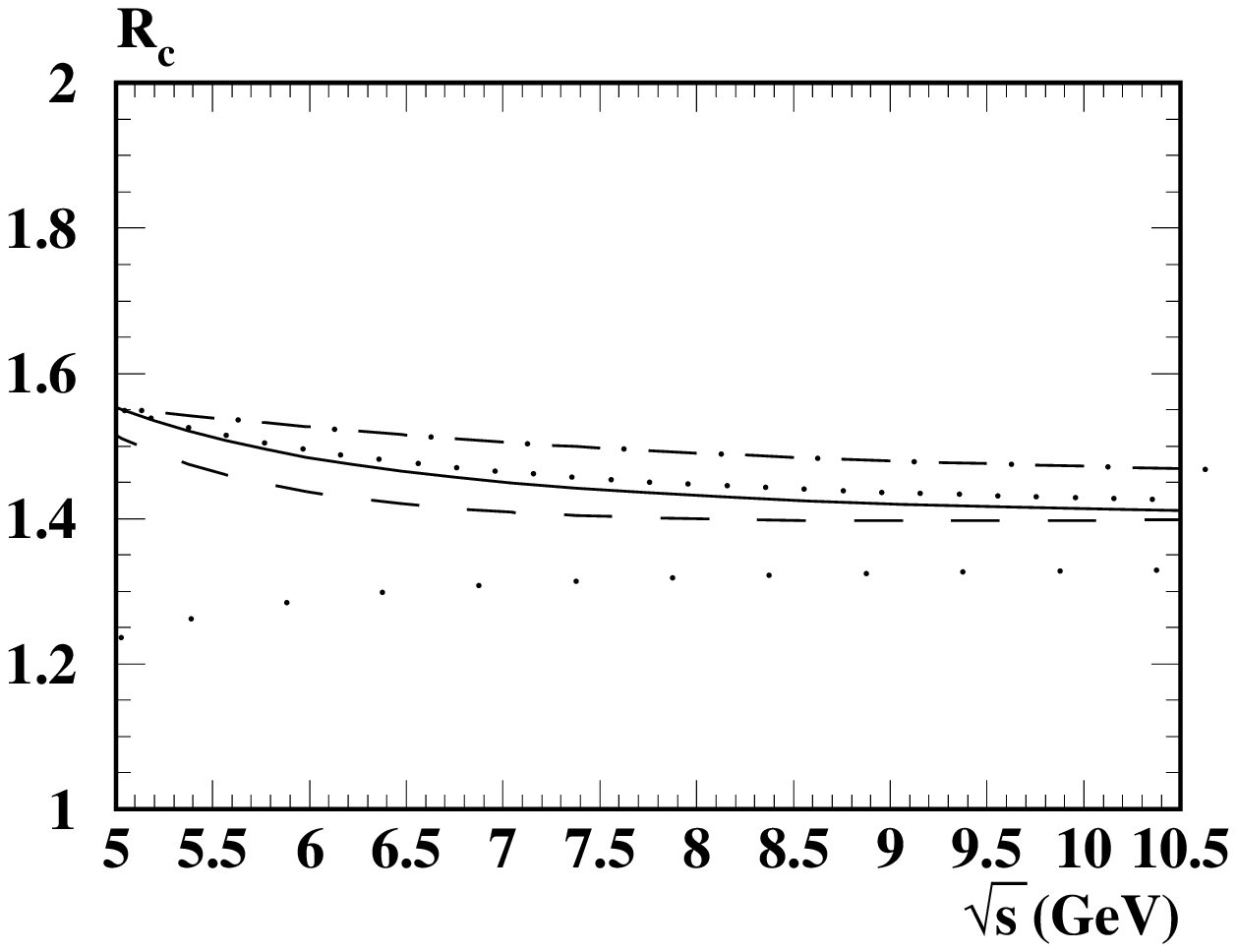}
&
\epsfxsize=7.0cm
\epsffile[100 280 460 560]{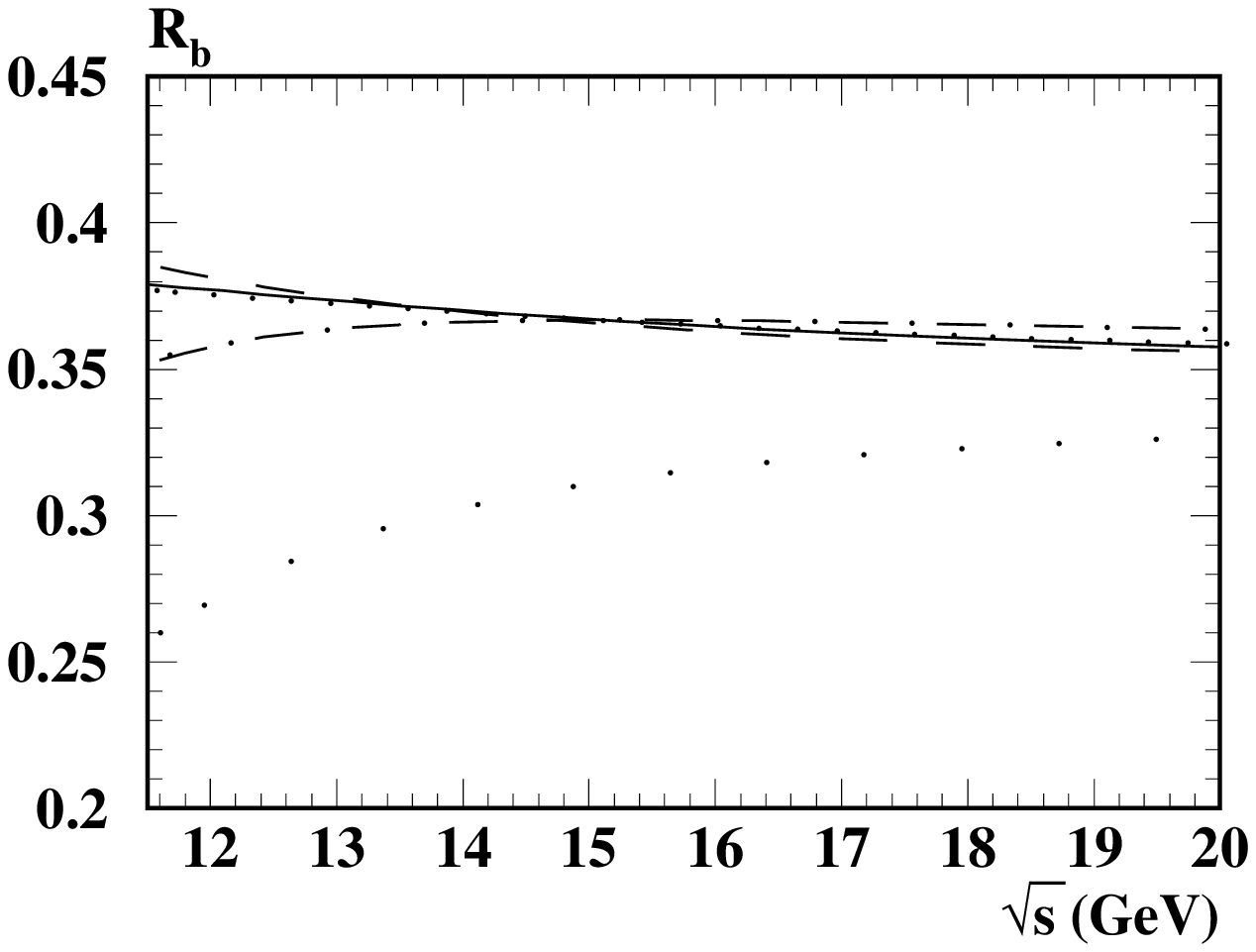}
\\
\epsfxsize=7.0cm
\epsffile[100 280 460 560]{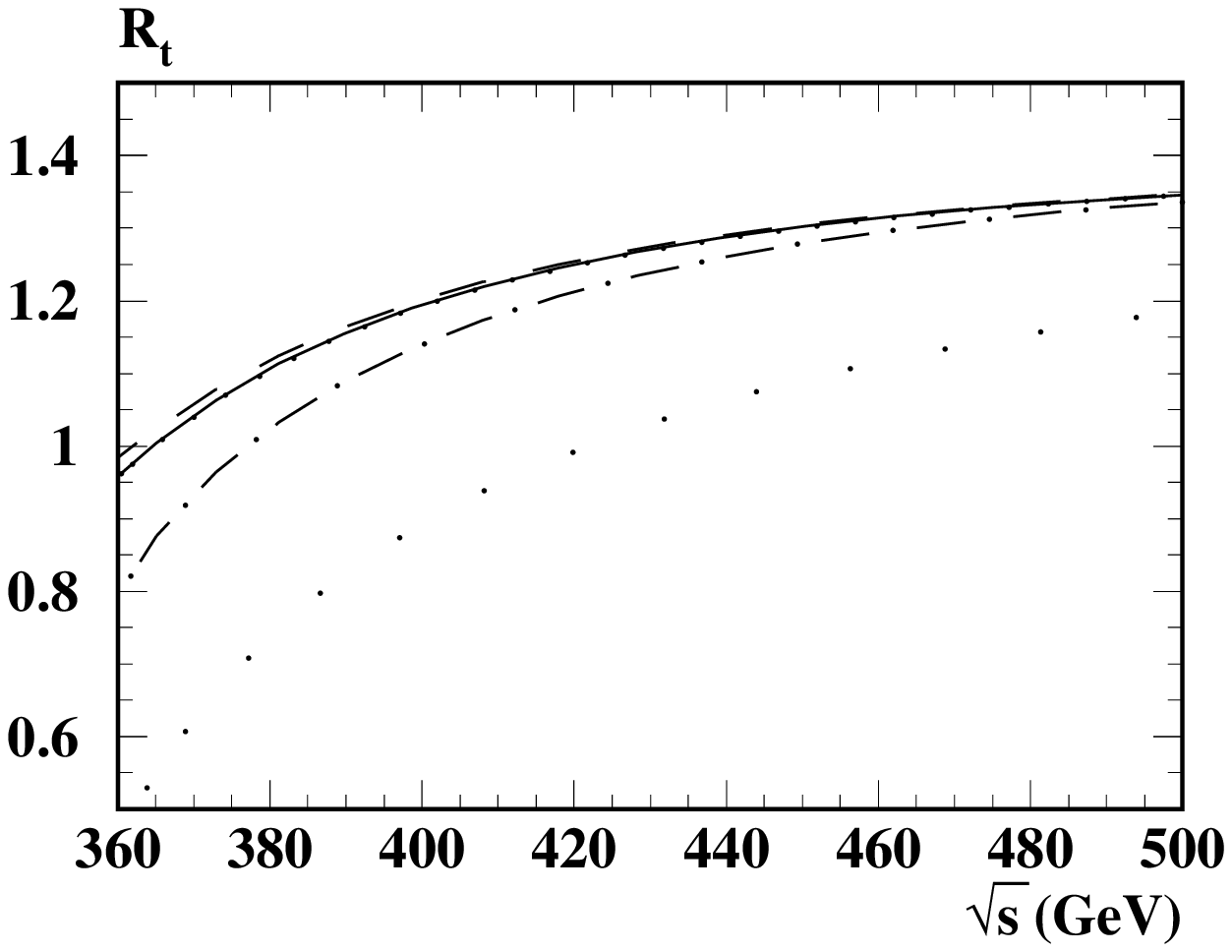}
\end{tabular}
\parbox{14.cm}{\small
\caption[]{\label{figrcbt} 
           The functions $R_c$, $R_b$ and $R_t$ in NLO plus dominant
           NNLO terms versus $\protect\sqrt{s}$ for three different scales,
           $\mu^2=M_Q^2$ (dashed), 
           $\mu^2=(2M_Q)^2$ (solid) and $\mu^2=s$ (dotted curves).
           For comparison, also shown are the 
           Born (wide dots)
           and ${\cal O}(\alpha_s)$ results
           ($\mu^2=(2M_Q)^2$, dash-dotted).
}}
\end{center}
\end{figure}

The collection of the results presented in the previous section
provides all tools necessary for a complete description of the 
cross section in NLO, including charm, bottom and top mass terms. 
This allows for the prediction of the charm, bottom and top
cross sections in the regions where quark masses cannot be neglected but 
where the resummation of Coulomb terms, characteristic for the regime 
very close to threshold, is not yet necessary, see the discussion below.
As stated above, the terms proportional to $Q_c^2$ and 
$\sum_{i=u, d, s} Q_i^2$ are invariant under renormalization group 
transformations separately, and only terms proportional to $Q_c^2$ will be 
considered in the following. 
Mutatis mutandis the same formulae are applicable to bottom quark
production. For top quarks only the piece induced by the electromagnetic
current will be considered, the axial part has been calculated 
in~\cite{CheKueSte97Pade,HoaTeu97,HarSte97}.

Fixed order perturbation theory is inapplicable very close to the 
production threshold, in the region where $\alpha_s/\beta$ is of order 
one or larger. In this region terms proportional to $(\alpha_s/\beta)^n$ 
have to be resummed. However, it will be demonstrated in Section~\ref{secthr}
that the first three terms in the 
perturbative expansion provide a good description down to fairly small 
values of $\beta$. Specifically, the relative deviation amounts to 
0.9/1.7/3.3\% for $C_F\pi\alpha_s/\beta = 2/2.5/\pi$, respectively. These 
values lie well within the radius of convergence 
$C_F\pi\alpha_s/\beta < 2\pi$ of the resummed series. 
Taking the requirement $C_F\pi\alpha_s/\beta < 2$ as a guiding 
principle and incorporating the running of the coupling constant one 
would admit the strictly perturbative, fixed order treatment down to 
energy values which are 1~GeV above the nominal threshold for bottom 
quarks and even less for charm quarks. However, since the perturbative 
treatment can only be applied beyond the highest $c\bar c$ and 
$b \bar b$ bound states, we take 5~GeV for charm and 11.5~GeV for bottom 
quarks as lowest center of mass energy values. For top, on the other hand, 
the limit $C_F\pi\alpha_s/\beta < 2$ 
corresponds to energies about $12$~GeV above $2M_t$.

\begin{figure}[t]
\begin{center}
\begin{tabular}{cc}
\leavevmode
\epsfxsize=6.5cm
\epsffile[100 280 460 560]{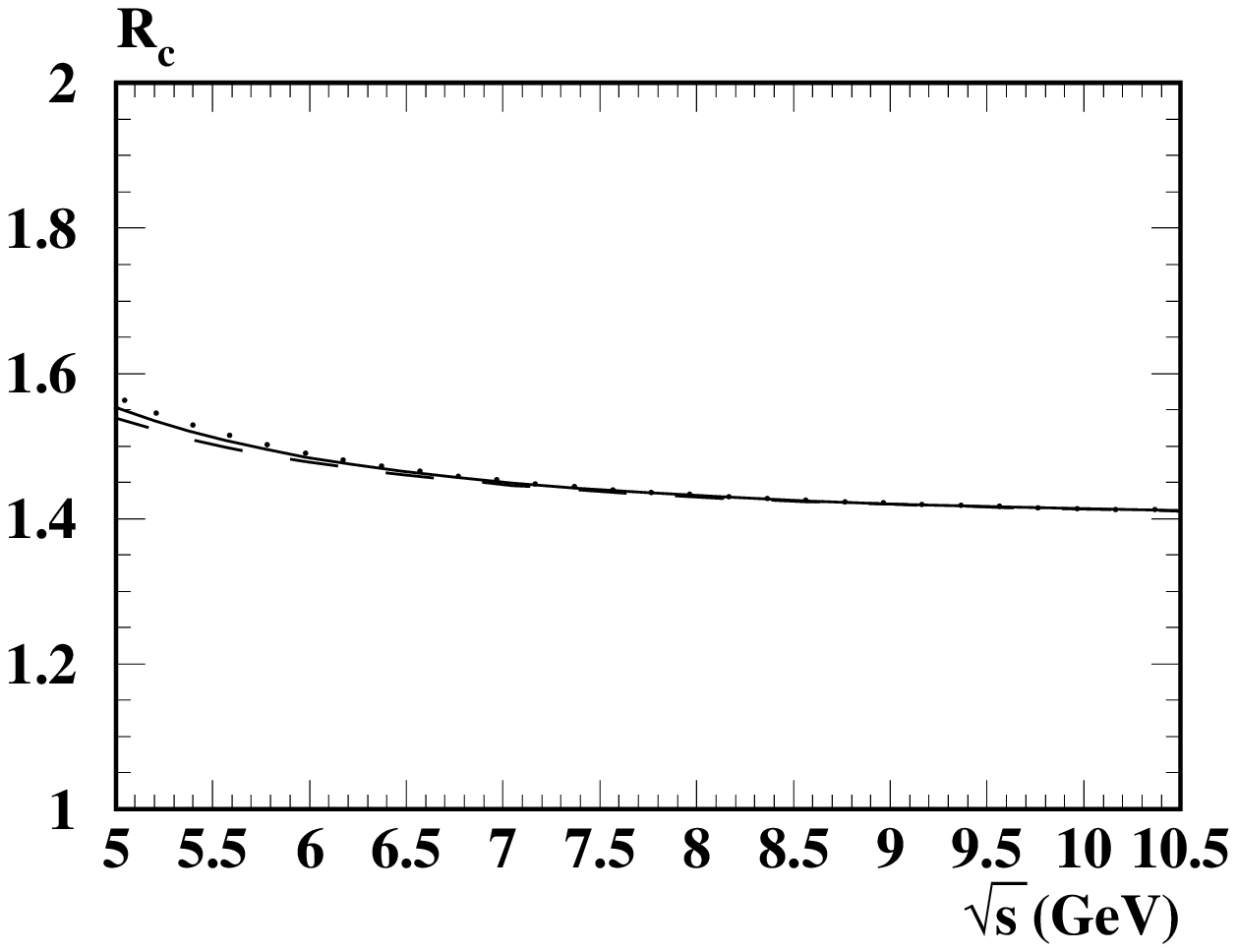}
&
\epsfxsize=6.5cm
\epsffile[100 280 460 560]{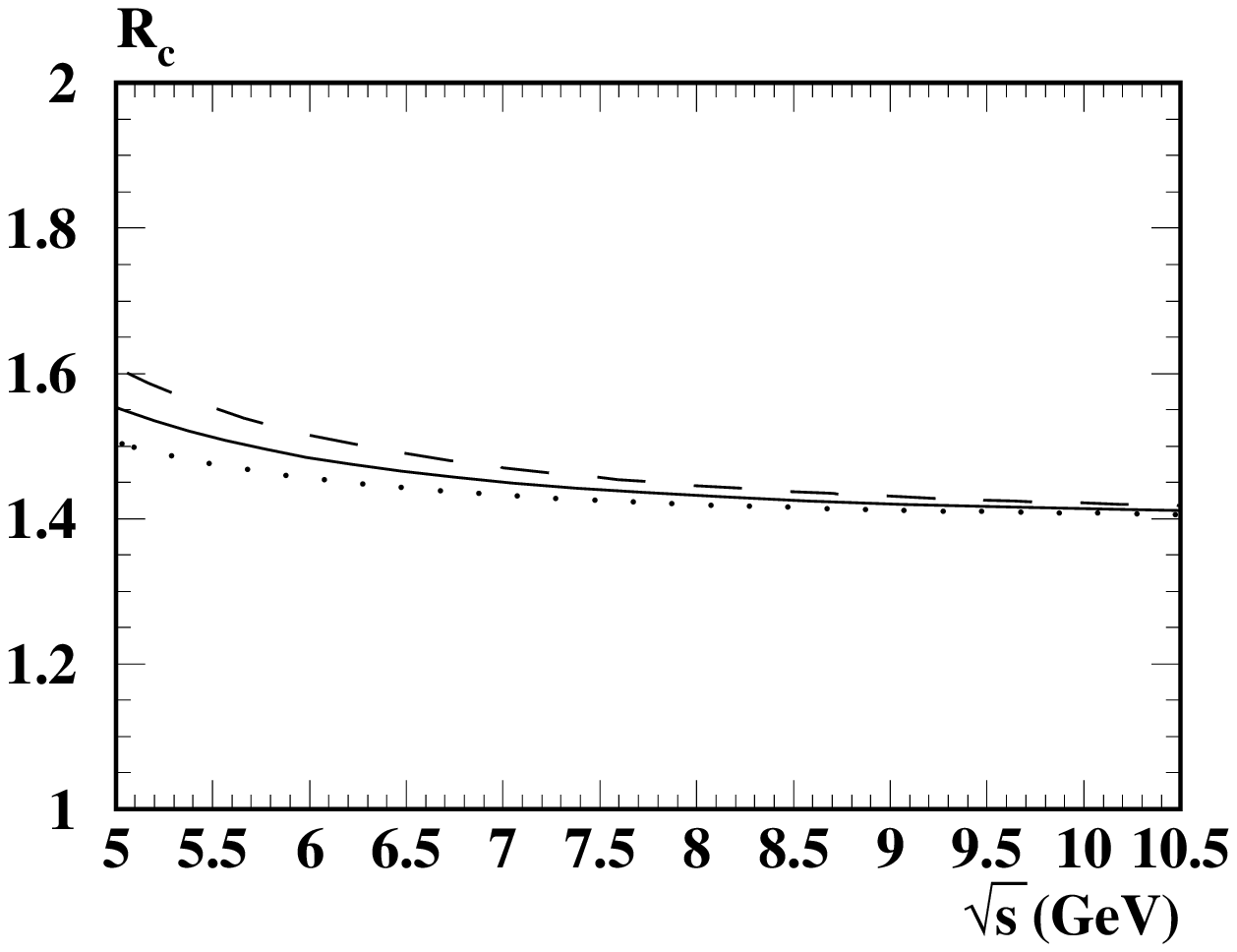}
\\
\epsfxsize=6.5cm
\epsffile[100 280 460 560]{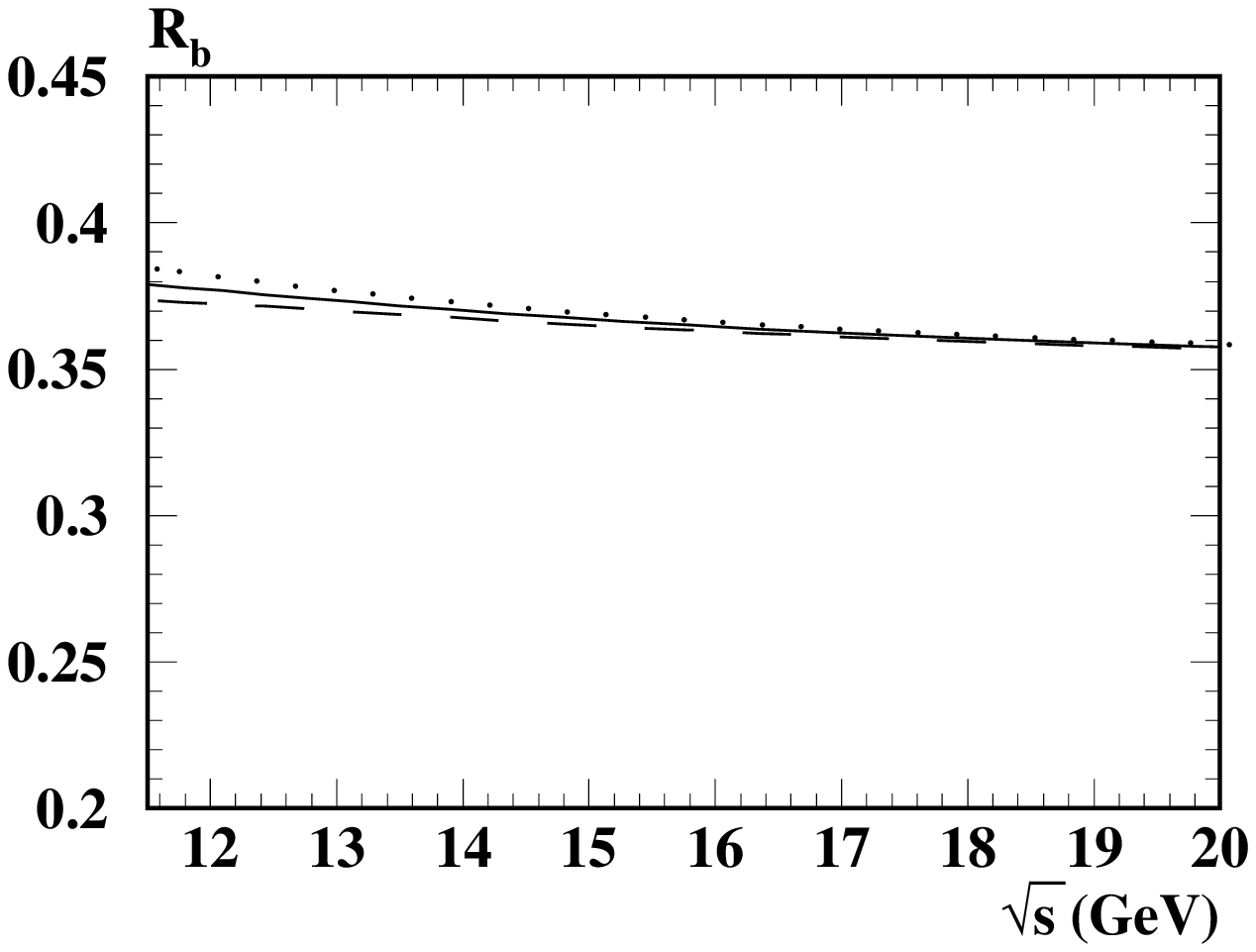}
&
\epsfxsize=6.5cm
\epsffile[100 280 460 560]{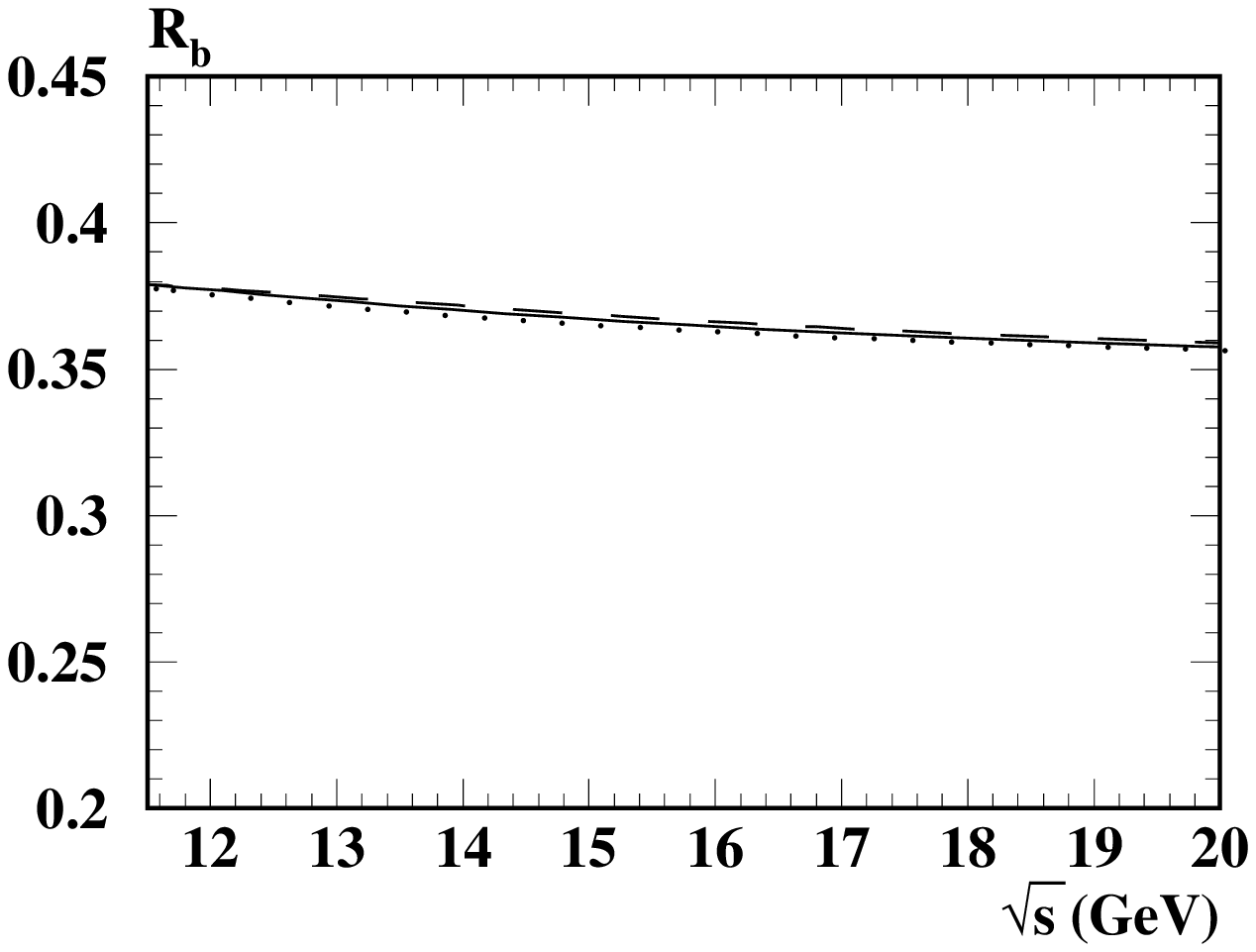}
\\
\epsfxsize=6.5cm
\epsffile[100 280 460 560]{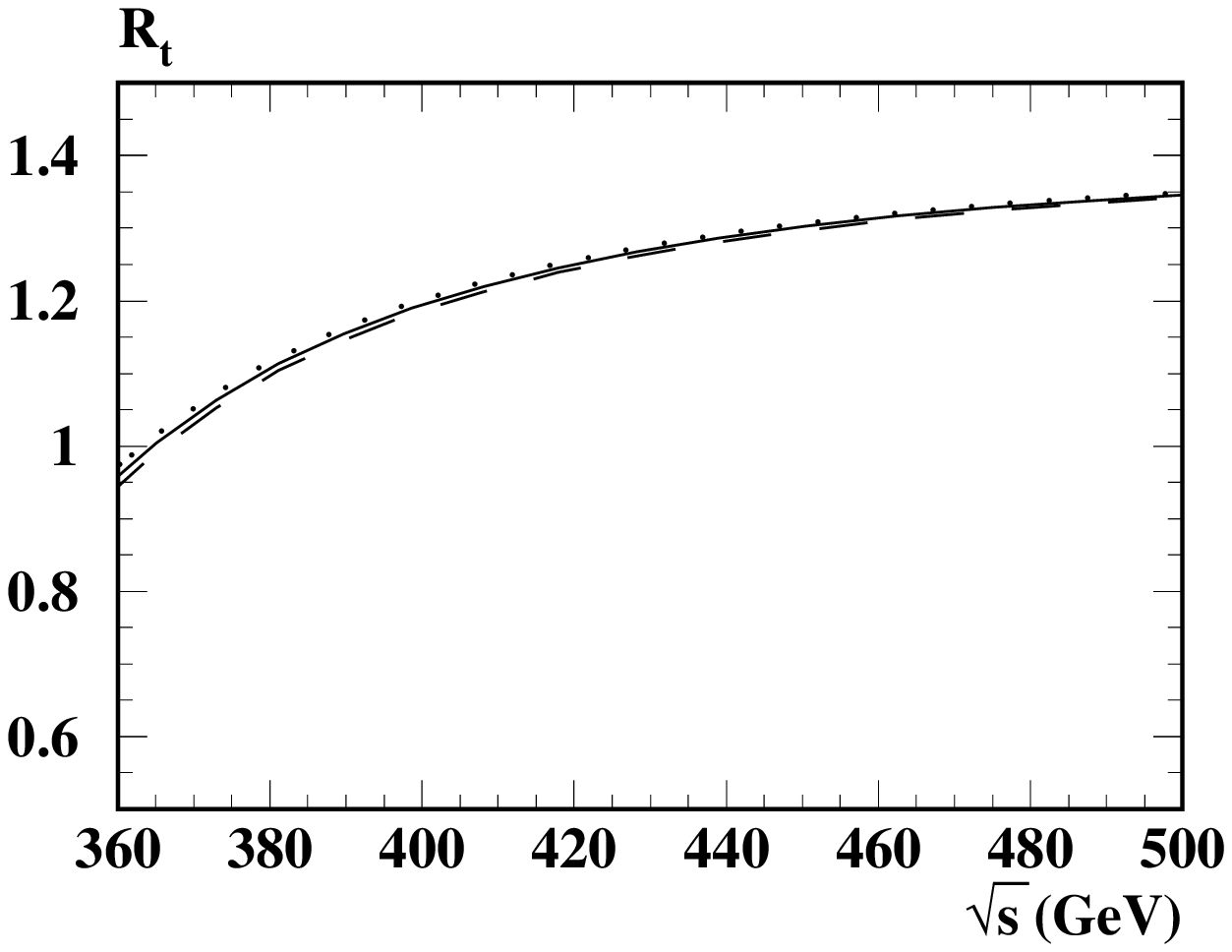}
&
\epsfxsize=6.5cm
\epsffile[100 280 460 560]{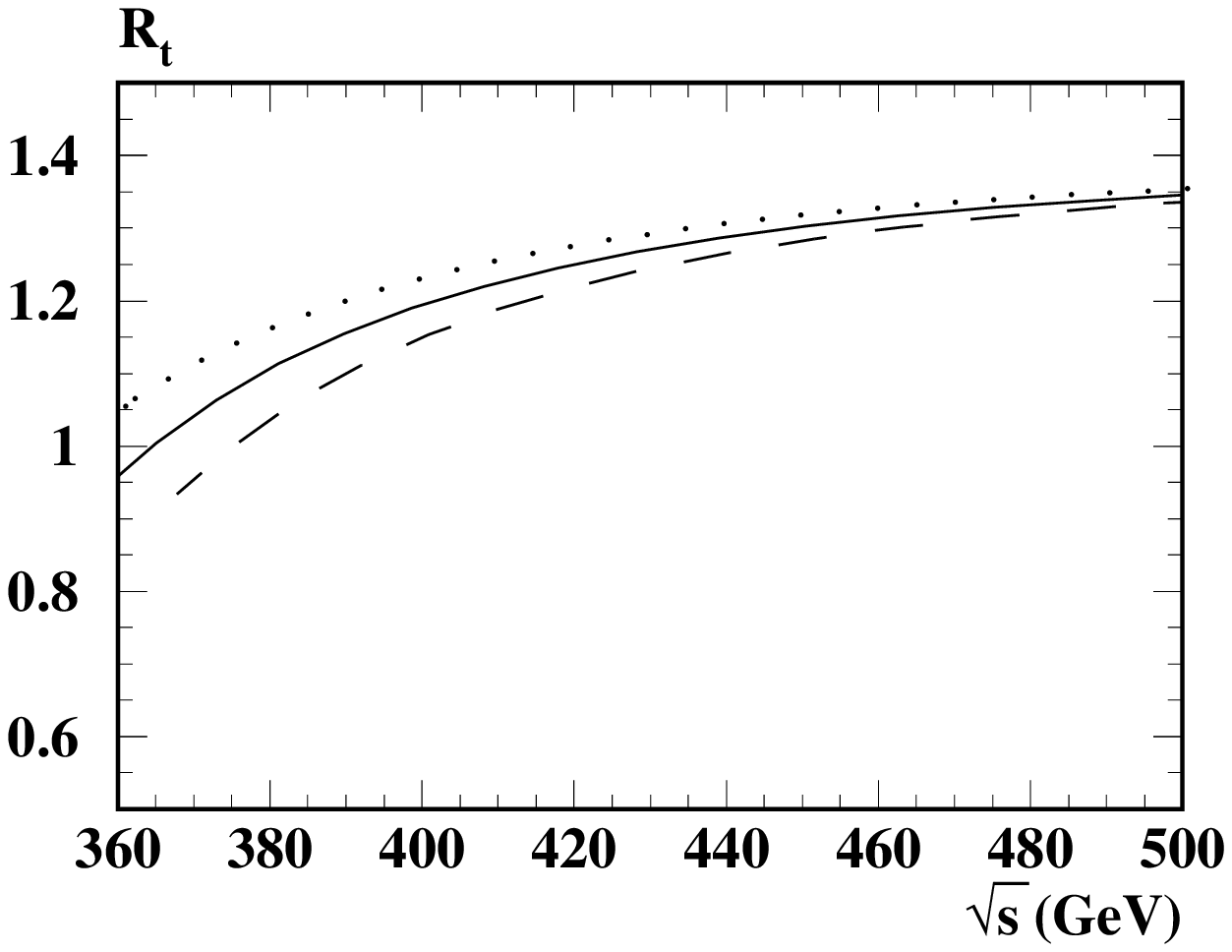}
\\
(a)
&
(b)
\end{tabular}
\parbox{14.cm}{\small
\caption[]{\label{figrcbtalsmass} Variation of $\alpha_s^{(5)}(M_Z^2)$ (a)
           and the quark masses (b). 
           In (a) the solid, dashed and dotted curves correspond
           to $\alpha_s^{(5)}(M_Z^2)=0.118,0.115$ and $0.121$, respectively.
           In Figs.~(b) for the solid curves  
           ($M_c$,$M_b$,$M_t$)=$(1.6,4.7,175)$~GeV
           is chosen.
           The dashed curves correspond to the upper ($(1.8,5.0,180)$~GeV)
           and the dotted curves to the lower limits ($(1.4,4.4,170)$~GeV).
           The scale $\mu^2=(2M_Q)^2$ has been adopted.
}}
\end{center}
\end{figure}

\begin{figure}[t]
\begin{center}
\begin{tabular}{c}
\leavevmode
\epsfxsize=15.5cm
\epsffile[100 280 460 560]{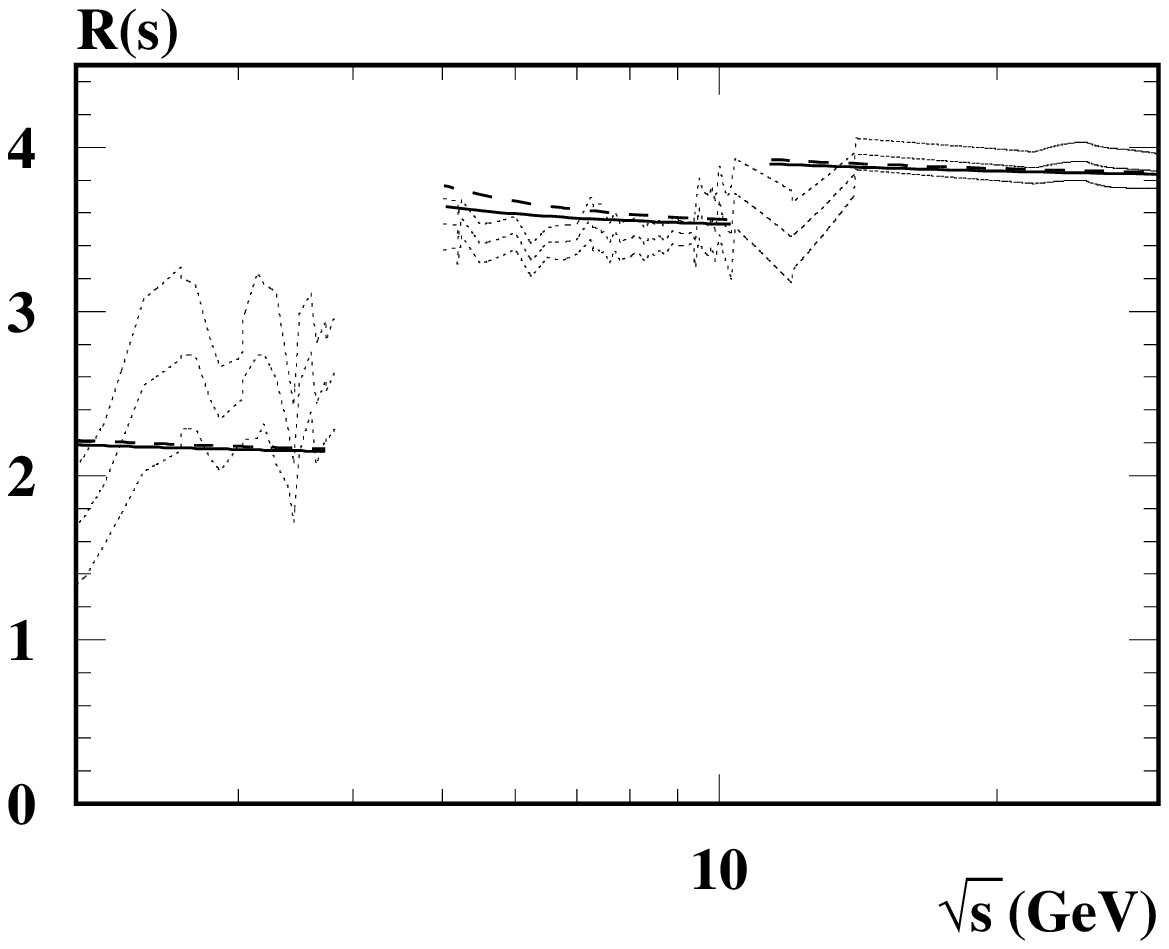}
\end{tabular}
\parbox{14.cm}{\small
\caption[]{\label{figrtot}
           $R(s)$ plotted against $\protect\sqrt{s}$.
           The scale $\mu^2=s$ has been adopted. The dashed curves 
           correspond to the values $M_c=1.8$~GeV, $M_b=5.0$~GeV
           and $\alpha_s(M_Z^2)=0.121$, whereas for the solid
           curves $M_c=1.4$~GeV, $M_b=4.4$~GeV
           and $\alpha_s(M_Z^2)=0.115$ is used.
           The dotted lines show a recent compilation of the 
           available experimental data. The central curves correspond 
           to the mean values, upper and lower curves indicate the 
           combined statistical and systematical errors. 
}}
\end{center}
\end{figure}

The compensation between
phase space suppression and Coulomb enhancement leads to a fairly flat
energy dependence of $R(s)$ even relatively close to threshold
(see Fig.~\ref{figrcbt}). In~\cite{CheKue94,CheHarKueSte97} 
it has been demonstrated for
${\cal O}(\alpha_s)$ and ${\cal O}(\alpha_s^2)$, respectively, that 
this behaviour is well approximated by the first terms in the 
large momentum expansion. Assuming
that the same line of reasoning is applicable also in order $\alpha_s^3$,
the dominant NNLO corrections can be 
incorporated~\cite{GorKatLar91SurSam91,CheKue90,CheKue97}:
\begin{eqnarray}
R_{c,ns}^{(3)} &=& 3\,\Bigg\{
-6.637 
+ 17.296\ln\frac{\mu^2}{s} 
+ 7.563\ln^2\frac{\mu^2}{s} 
\nonumber\\&&\mbox{}
+ n_f\left(
-1.200 
- 2.088\ln\frac{\mu^2}{s} 
- 0.917\ln^2\frac{\mu^2}{s}
\right)
\nonumber\\&&\mbox{}
+n_f^2\left( 
-0.005 
+ 0.038\ln\frac{\mu^2}{s} 
+ 0.028\ln^2\frac{\mu^2}{s}
\right) 
+ \frac{M_c^2}{s}\Bigg[
347.168 
- 378.000\ln\frac{\mu^2}{M_c^2} 
\nonumber\\&&\mbox{}
- 9.000\ln^2\frac{\mu^2}{M_c^2} 
+ 974.250\ln\frac{\mu^2}{s} 
- 114.000\ln\frac{\mu^2}{M_c^2}\ln\frac{\mu^2}{s} 
+ 213.750\ln^2\frac{\mu^2}{s} 
\nonumber\\&&\mbox{}
+ n_f\left(
-67.619 
+ 17.333\ln\frac{\mu^2}{M_c^2} 
+ 2.000\ln^2\frac{\mu^2}{M_c^2} 
- 82.167\ln\frac{\mu^2}{s} 
\right.\nonumber\\&&\left.\mbox{}
+ 4.000\ln\frac{\mu^2}{M_c^2}\ln\frac{\mu^2}{s} 
- 17.000\ln^2\frac{\mu^2}{s}
\right) 
\nonumber\\&&\mbox{}
+ n_f^2\left(
1.218 
+ 1.444\ln\frac{\mu^2}{s} 
+ 0.333\ln^2\frac{\mu^2}{s}
\right)
\Bigg]
\Bigg\}.
\label{eqrthirdorder}
\end{eqnarray}
Inclusion of these terms will lead to the correct NNLO predictions for
larger energies, say above $7$ or $8$~GeV, allowing at the same time for
a smooth transition to NLO accuracy for lower energies.
The singlet terms proportional to $(\sum Q_i)^2$ are 
small~\cite{GorKatLar91SurSam91}
and have been neglected in Eq.~(\ref{eqrthirdorder}).
In total one thus finds:
\begin{eqnarray}
R_c &=& Q_c^2\left(R_c^{(0)} + \frac{\alpha_s^{(4)}(\mu^2)}{\pi} C_FR_c^{(1)}
+
\left(\frac{\alpha_s^{(4)}(\mu^2)}{\pi}\right)^2 
R_c^{(2)}
+
\left(\frac{\alpha_s^{(4)}(\mu^2)}{\pi}\right)^3\,
R_{c,ns}^{(3)}
\right),
\label{eqrctot}
\end{eqnarray}
where $R_{c,ns}^{(3)}$ denotes the non-singlet contribution at 
${\cal O}(\alpha_s^3)$. In Fig.~\ref{figrcbt} the predictions
are shown for the charm, bottom and top cross sections in the energy 
regions discussed above. For the value of the coupling 
$\alpha_s^{(5)}(M_Z^2) = 0.118$ has been adopted
and the on-shell quark masses have been chosen to be
$M_c=1.6$~GeV, $M_b=4.7$~GeV and $M_t=175$~GeV. 
(Note that a complete prediction of the top quark cross section 
would require the incorporation of the axial
contribution.)
In order to study the 
sensitivity of the results to the renormalization scale, $\mu^2$ has been 
chosen as $M_Q^2$ ($Q=c,b,t$; dashed curves), 
$4M_Q^2$ (solid curves) and $s$ (dotted 
curves). The results are relatively stable against this variation. 
This agreement, despite the appearance of large logarithms $\ln s/M_Q^2$
for $\mu^2=M_Q^2$, is a consequence of the NNLO approximation valid
at the high energy end. The largest sensitivity towards the choice of 
$\mu^2$ is observed in the intermediate and low energy region for charm 
production, where $\alpha_s$ is large and the corrections are enhanced by 
large contributions proportional to $\ln\beta$ and $1/\beta$.
For comparison $R_c$, $R_b$ and $R_t$ are also plotted in the
Born (wide dots)
and leading order approximation (dash-dotted, $\mu^2=(2M_Q)^2$). 
The cubic corrections in $\alpha_s$ are rather small so that we 
do not show the NLO corrections separately.
For the case of the charm and bottom quark the ``remainder''
of the Coulomb singularity is still visible and both $R_c$ and $R_b$
raise for $\sqrt{s}\to 5$~GeV and $\sqrt{s}\to 11.5$~GeV, respectively.
For the top quark, however, even $10$~GeV above the nominal
pair production threshold a singular behaviour is not visible and $R_t$
decreases as $\sqrt{s}$ approaches $360$~GeV.

To study the dependence on the input parameters $M_Q$ and 
$\alpha_s^{(5)}(M_Z^2)$, we adopt 
the choice $\mu^2 = 4M_Q^2$ and vary $\alpha_s^{(5)}(M_Z^2)$ 
within the currently 
quoted range $\pm 0.03$ (see Fig.~\ref{figrcbtalsmass}(a)) 
and the quark masses 
in the range as indicated in Fig.~\ref{figrcbtalsmass}(b). These changes 
indicate the present uncertainties.
For $R_c$ and $R_b$ the higher values
of the quark masses also lead to 
larger cross sections --- a consequence of the enhanced Coulomb forces
for fixed $\sqrt{s}$. For 
the top quark the phase space effect is still dominant and the
cross section for the choice $M_t=180$~GeV is smaller than for $M_t=170$~GeV.

\begin{table}[t]
\renewcommand{\arraystretch}{1.3}
\begin{center}
{\scriptsize
\begin{tabular}{|l|l||l||r|r|r|r|r|r|r|}
\hline\hline 
$\alpha_s^{(5)}(M_Z^2)$ & $M_c$ & $\sqrt{s}$ 
&$5$&$6$&$7$&$8$&$9$&$9.98$&$10.52$
\\ 
 & (GeV) & (GeV)  &  &  &  &  &  &  & \\ \hline\hline
$0.115$ & $1.4$ & $R_c(s)$
&$  1.505$&$  1.467$&$  1.447$&$  1.434$&$  1.426$&$  1.420$&$  1.418$
\\ 
$0.115$ & $1.4$ & $R(s)$
&$  3.641$&$  3.596$&$  3.570$&$  3.554$&$  3.542$&$  3.534$&$  3.530$
\\ \hline
$0.115$ & $1.6$ & $R_c(s)$
&$  1.538$&$  1.486$&$  1.459$&$  1.443$&$  1.432$&$  1.425$&$  1.422$
\\ 
$0.115$ & $1.6$ & $R(s)$
&$  3.674$&$  3.615$&$  3.582$&$  3.562$&$  3.548$&$  3.538$&$  3.534$
\\ \hline
$0.115$ & $1.8$ & $R_c(s)$
&$  1.575$&$  1.507$&$  1.473$&$  1.452$&$  1.439$&$  1.431$&$  1.427$
\\ 
$0.115$ & $1.8$ & $R(s)$
&$  3.711$&$  3.636$&$  3.596$&$  3.572$&$  3.556$&$  3.544$&$  3.539$
\\ \hline
\hline
$0.118$ & $1.4$ & $R_c(s)$
&$  1.516$&$  1.473$&$  1.451$&$  1.438$&$  1.430$&$  1.424$&$  1.421$
\\ 
$0.118$ & $1.4$ & $R(s)$
&$  3.659$&$  3.608$&$  3.581$&$  3.563$&$  3.550$&$  3.541$&$  3.537$
\\ \hline
$0.118$ & $1.6$ & $R_c(s)$
&$  1.553$&$  1.494$&$  1.465$&$  1.447$&$  1.436$&$  1.429$&$  1.425$
\\ 
$0.118$ & $1.6$ & $R(s)$
&$  3.696$&$  3.629$&$  3.594$&$  3.572$&$  3.557$&$  3.546$&$  3.542$
\\ \hline
$0.118$ & $1.8$ & $R_c(s)$
&$  1.598$&$  1.519$&$  1.480$&$  1.458$&$  1.444$&$  1.434$&$  1.430$
\\ 
$0.118$ & $1.8$ & $R(s)$
&$  3.741$&$  3.654$&$  3.609$&$  3.583$&$  3.565$&$  3.552$&$  3.547$
\\ \hline
\hline
$0.121$ & $1.4$ & $R_c(s)$
&$  1.527$&$  1.480$&$  1.456$&$  1.442$&$  1.433$&$  1.427$&$  1.424$
\\ 
$0.121$ & $1.4$ & $R(s)$
&$  3.677$&$  3.621$&$  3.591$&$  3.572$&$  3.559$&$  3.549$&$  3.545$
\\ \hline
$0.121$ & $1.6$ & $R_c(s)$
&$  1.569$&$  1.503$&$  1.471$&$  1.452$&$  1.440$&$  1.432$&$  1.429$
\\ 
$0.121$ & $1.6$ & $R(s)$
&$  3.719$&$  3.644$&$  3.605$&$  3.582$&$  3.566$&$  3.554$&$  3.549$
\\ \hline
$0.121$ & $1.8$ & $R_c(s)$
&$  1.622$&$  1.531$&$  1.488$&$  1.463$&$  1.448$&$  1.438$&$  1.434$
\\ 
$0.121$ & $1.8$ & $R(s)$
&$  3.771$&$  3.672$&$  3.623$&$  3.593$&$  3.574$&$  3.561$&$  3.555$
\\ \hline
\hline
\end{tabular}
}
\parbox{14.cm}{\small
\caption{\label{tabrs1}Numerical values for $R(s)$ and $R_c(s)$.
The scale $\mu^2=s$ has been chosen. For the evaluation of 
$R_c(s)$ Eq.~(\ref{eqrctot}) has been used. $R(s)$ is the proper sum of
$R_{uds}(s)$ (Eq.~(\ref{eqruds})) and $R_c(s)$ with $M_b=4.7$~GeV.
}}
\end{center}
\end{table}

\begin{table}[t]
\renewcommand{\arraystretch}{1.3}
\begin{center}
{\scriptsize
\begin{tabular}{|l|l||l||r|r|r|r|r|r|}
\hline\hline 
$\alpha_s^{(5)}(M_Z^2)$ & $M_b$ & $\sqrt{s}$ 
&$11.5$&$12$&$13$&$14$&$15$&$40$
\\ 
 & (GeV) & (GeV)  &  &  &  &  &  & \\ \hline\hline
$0.115$ & $4.4$ & $R_b(s)$
&$  0.372$&$  0.371$&$  0.368$&$  0.366$&$  0.364$&$  0.349$
\\ 
$0.115$ & $4.4$ & $R(s)$
&$  3.899$&$  3.895$&$  3.888$&$  3.881$&$  3.876$&$  3.827$
\\ \hline
$0.115$ & $4.7$ & $R_b(s)$
&$  0.372$&$  0.371$&$  0.369$&$  0.367$&$  0.365$&$  0.349$
\\ 
$0.115$ & $4.7$ & $R(s)$
&$  3.899$&$  3.895$&$  3.889$&$  3.882$&$  3.877$&$  3.827$
\\ \hline
$0.115$ & $5.0$ & $R_b(s)$
&$  0.370$&$  0.370$&$  0.369$&$  0.368$&$  0.366$&$  0.350$
\\ 
$0.115$ & $5.0$ & $R(s)$
&$  3.897$&$  3.894$&$  3.889$&$  3.883$&$  3.878$&$  3.827$
\\ \hline
\hline
$0.118$ & $4.4$ & $R_b(s)$
&$  0.376$&$  0.374$&$  0.371$&$  0.368$&$  0.365$&$  0.350$
\\ 
$0.118$ & $4.4$ & $R(s)$
&$  3.911$&$  3.906$&$  3.897$&$  3.890$&$  3.884$&$  3.832$
\\ \hline
$0.118$ & $4.7$ & $R_b(s)$
&$  0.377$&$  0.375$&$  0.372$&$  0.370$&$  0.367$&$  0.350$
\\ 
$0.118$ & $4.7$ & $R(s)$
&$  3.911$&$  3.907$&$  3.899$&$  3.891$&$  3.885$&$  3.832$
\\ \hline
$0.118$ & $5.0$ & $R_b(s)$
&$  0.377$&$  0.376$&$  0.373$&$  0.371$&$  0.368$&$  0.350$
\\ 
$0.118$ & $5.0$ & $R(s)$
&$  3.911$&$  3.907$&$  3.900$&$  3.893$&$  3.887$&$  3.832$
\\ \hline
\hline
$0.121$ & $4.4$ & $R_b(s)$
&$  0.381$&$  0.378$&$  0.374$&$  0.370$&$  0.367$&$  0.350$
\\ 
$0.121$ & $4.4$ & $R(s)$
&$  3.922$&$  3.917$&$  3.907$&$  3.899$&$  3.892$&$  3.837$
\\ \hline
$0.121$ & $4.7$ & $R_b(s)$
&$  0.383$&$  0.380$&$  0.376$&$  0.372$&$  0.369$&$  0.350$
\\ 
$0.121$ & $4.7$ & $R(s)$
&$  3.924$&$  3.919$&$  3.909$&$  3.901$&$  3.894$&$  3.837$
\\ \hline
$0.121$ & $5.0$ & $R_b(s)$
&$  0.384$&$  0.382$&$  0.377$&$  0.374$&$  0.371$&$  0.351$
\\ 
$0.121$ & $5.0$ & $R(s)$
&$  3.926$&$  3.920$&$  3.911$&$  3.903$&$  3.896$&$  3.837$
\\ \hline
\hline
\end{tabular}
}
\parbox{14.cm}{\small
\caption{\label{tabrs2}Numerical values for $R(s)$ and $R_b(s)$.
The scale $\mu^2=s$ has been chosen. For the evaluation of 
$R_b(s)$ Eq.~(\ref{eqrctot}) with obvious modifications has been used. 
$R(s)$ is the proper sum of
$R_{udsc}(s)$ (obtained from Eq.~(\ref{eqruds}))
where in addition charm mass effects of order $M_c^2/s$ have been 
included 
(Eqs.~(\ref{eqas012m}) and (\ref{eqas3add}))
and $R_b(s)$ ($M_c=1.6$~GeV).
}}
\end{center}
\end{table}

\begin{table}[t]
\renewcommand{\arraystretch}{1.3}
\begin{center}
{\scriptsize
\begin{tabular}{|l||l||r|r|r|r|r|r|r|}
\hline\hline 
$\mu$ & $\sqrt{s}$ 
&$5$&$6$&$7$&$8$&$9$&$9.98$&$10.52$
\\ 
 & (GeV)  &  &  &  &  &  &  & \\ \hline\hline
$M_c$ & $R_c(s)$
&$  1.514$&$  1.437$&$  1.409$&$  1.400$&$  1.397$&$  1.398$&$  1.399$
\\ \hline
$2M_c$ & $R_c(s)$
&$  1.553$&$  1.484$&$  1.450$&$  1.432$&$  1.421$&$  1.414$&$  1.411$
\\ \hline
$\sqrt{s}$ & $R_c(s)$
&$  1.553$&$  1.494$&$  1.465$&$  1.447$&$  1.436$&$  1.429$&$  1.425$
\\ \hline
$2\sqrt{s}$ & $R_c(s)$
&$  1.541$&$  1.493$&$  1.466$&$  1.449$&$  1.438$&$  1.430$&$  1.427$
\\ \hline
\hline
\end{tabular}
}
\parbox{14.cm}{\small
\caption{\label{tabrs1mu}
         Numerical values for $R_c(s)$ for different 
         choices of $\protect\sqrt{s}$ and $\mu$.
         The values $\alpha_s^{(5)}(M_Z^2)=0.118$, $M_c=1.6$~GeV 
         and $M_b=4.7$~GeV have been chosen.
}}
\end{center}
\end{table}

\begin{table}[t]
\renewcommand{\arraystretch}{1.3}
\begin{center}
{\scriptsize
\begin{tabular}{|l||l||r|r|r|r|r|r|r|}
\hline\hline 
$\mu$ & $\sqrt{s}$ 
&$11.5$&$12$&$13$&$14$&$15$&$40$
\\ 
 & (GeV)  &  &  &  &  &  & \\ \hline\hline
$M_b$ & $R_b(s)$
&$  0.386$&$  0.381$&$  0.375$&$  0.370$&$  0.366$&$  0.349$
\\ \hline
$2M_b$ & $R_b(s)$
&$  0.379$&$  0.377$&$  0.373$&$  0.370$&$  0.367$&$  0.349$
\\ \hline
$\sqrt{s}$ & $R_b(s)$
&$  0.377$&$  0.375$&$  0.372$&$  0.370$&$  0.367$&$  0.350$
\\ \hline
$2\sqrt{s}$ & $R_b(s)$
&$  0.370$&$  0.370$&$  0.369$&$  0.368$&$  0.366$&$  0.350$
\\ \hline
\hline
\end{tabular}
}
\parbox{14.cm}{\small
\caption{\label{tabrs2mu}
         Numerical values for $R_b(s)$ for different 
         choices of $\protect\sqrt{s}$ and $\mu$.
         The values $\alpha_s^{(5)}(M_Z^2)=0.118$, $M_c=1.6$~GeV 
         and $M_b=4.7$~GeV have been chosen.
}}
\end{center}
\end{table}

It is now possible to give a prediction for $R(s)$ in the energy range 
above $\sqrt{s}\approx2$~GeV with the exception of small windows of
$1$ to $2$~GeV above the thresholds for open charm and bottom production,
respectively.
Below the charm meson threshold at $\sqrt{s}=2M_D$ the charm quark
is treated as 
heavy and Eqs.~(\ref{eqrmassless}) and (\ref{eqrqb}) are used with
$n_\ell=3$, and with $\alpha_s^{(4)}$ and $M_b$ replaced by
$\alpha_s^{(3)}$ and $M_c$, respectively.
For $5$~GeV$\lsim\sqrt{s}\lsim10.5$~GeV
Eqs.~(\ref{eqruds}) and (\ref{eqdecomprtwo})
are directly applicable with $n_f=n_\ell+1=4$. 
For the case with external charm and internal bottom quark an 
expression analogue to Eq.~(\ref{eqrqb}) is used.
Above the bottom threshold the sum over $i$ in Eqs.~(\ref{eqrmassless})
and (\ref{eqr2qc}) includes also the charm contribution and consequently
$n_\ell=4$ has to be chosen,
$\alpha_s^{(4)}$ and $M_c$ have to be replaced by
$\alpha_s^{(5)}$ and $M_b$. 
The dominant charm mass terms are included through the leading terms
in the $M_c^2/s$ approximation.
The quadratic charm mass corrections of order $\alpha_s^3$ are given in
Eq.~(\ref{eqrthirdorder}). The corresponding terms up to 
order $\alpha_s^2$ read as follows~\cite{GorKatLar86}:
\begin{eqnarray}
\delta R_{M_c^2}^{(0+1+2)} &=& 3 Q_c^2 \frac{M_c^2}{s}
\Bigg[
12 \frac{\alpha_s^{(5)}(\mu^2)}{\pi}
+
\left(\frac{\alpha_s^{(5)}(\mu^2)}{\pi}\right)^2
\left(
\frac{189}{2} 
- 24\ln\frac{\mu^2}{M_c^2} 
- 57\ln\frac{s}{\mu^2}
\right.\nonumber\\&&\mbox{}\left.\qquad\qquad
+n_f\left(-\frac{13}{3} + 2\ln\frac{s}{\mu^2}\right)
\right)
\Bigg].
\label{eqas012m}
\end{eqnarray}
In addition the term
\begin{eqnarray}
3\,\frac{M_c^2}{s}\, \left( -7.877 + 0.350 n_f \right)
\label{eqas3add}
\end{eqnarray}
which comes from the expansion in $M_c^2/s$ of diagrams with 
internal charm loops in $b\bar{b}$ production~\cite{CheKue90,CheKue97}
has to be added to $R_{b,ns}^{(3)}$.

In Fig.~\ref{figrtot} $R(s)$ is plotted versus $\sqrt{s}$
for the three energy intervals. Solid and dashed curves show 
our prediction from perturbative QCD, where the extreme values 
from the variation of $\alpha_s^{(5)}(M_Z^2)$ and the masses 
are considered and the scale $\mu^2=s$ has been adopted.
For comparison we also plot a recent compilation of the 
available experimental data~\cite{data}
from the inclusive measurements of $R$. 
The central dotted curves correspond to the mean values, upper and 
lower dotted curves indicate the combined statistical and 
systematical errors.

In Tabs.~\ref{tabrs1} and \ref{tabrs2} numerical values are listed 
for the energy range between the charm and bottom threshold and 
above the bottom threshold, respectively.
For completeness the terms of order $\alpha_s^3$ to 
$R_{\rm light}$ are also taken into account. They are obtained from
Eq.~(\ref{eqrthirdorder}) neglecting the mass corrections with
$n_f=4$ (Tab.~\ref{tabrs1}) and $n_f=5$ (Tab.~\ref{tabrs2}), respectively.
In Tabs.~\ref{tabrs1mu} and \ref{tabrs2mu} the dependence of 
$R_c$ and $R_b$ on the renormalization scale
$\mu$ is displayed with $M_Q$ and $\alpha_s$ fixed to  their 
central values.

It is evident that the uncertainties in the prediction are far below the 
experimental errors in this ``low'' energy region. These results 
could therefore be used to fit the currently available data and to allow for
an improved input into the analysis of the running 
electromagnetic coupling constant $\alpha$.


\section{Hadronic vacuum polarization and the running electroweak coupling}
\label{secqed}

One of the important ingredients of electroweak
precision tests is the effect of the hadronic vacuum polarization on the
running of the electromagnetic coupling. Using a dispersion relation, it is
expressed~\cite{delal}
through $R(s)$
\begin{eqnarray}
\Delta\alpha_{\rm had}^{(5)}(M_Z^2)
&=&
-\frac{\alpha M_Z^2}{3\pi}\,\mbox{Re}\,
\int_{4m_\pi^2}^\infty\,{\rm d}
s\,\frac{R(s)}{s\left(s-M_Z^2-i\epsilon\right)}
\label{eqdelal}
\end{eqnarray}
and contributes together with the well known leptonic
contributions to the running of the electromagnetic coupling
\begin{eqnarray}
\alpha(s) &=& \frac{\alpha(0)}
      {1-\Delta\alpha^{(5)}_{\rm had}(s)-\Delta\alpha_{\rm lep}(s)},
\end{eqnarray}
where $\alpha(0)=1/137.0359895$ is the fine structure constant.
Top quark contributions are not considered in this context.
Also QED corrections are not included as they are of the
order of a few per mill only.

\begin{table}[t]
\renewcommand{\arraystretch}{1.3}
\begin{center}
{\scriptsize
\begin{tabular}{|l||r|r|r||r|r|r||r|r|r|}
\hline\hline 
   $\alpha_s^{(5)}(M_Z^2)$
 & \multicolumn{3}{|c||}{$0.115$}
 & \multicolumn{3}{|c||}{$0.118$}
 & \multicolumn{3}{|c|}{$0.121$} \\ \hline\hline
Energy range &&&&&&&&& \\ \hline
$  2.00-  2.50$~GeV
&\multicolumn{3}{|c||}{  7.54}
&\multicolumn{3}{|c||}{  7.58}
&\multicolumn{3}{|c|}{  7.62} \\ 
$  2.50-  3.00$~GeV
&\multicolumn{3}{|c||}{  6.12}
&\multicolumn{3}{|c||}{  6.15}
&\multicolumn{3}{|c|}{  6.18} \\ 
$  3.00-  3.73$~GeV
&\multicolumn{3}{|c||}{  7.27}
&\multicolumn{3}{|c||}{  7.30}
&\multicolumn{3}{|c|}{  7.33} \\ 
\hline\hline
$M_c$ (GeV) & $1.4$&$1.6$&$1.8$&$1.4$&$1.6$&$1.8$&$1.4$&$1.6$&$1.8$ \\ \hline
Energy range &&&&&&&&& \\ \hline
\hline
$  3.73-  5.00$~GeV
&  9.73&  9.74&  9.74&  9.77&  9.77&  9.77&  9.80&  9.80&  9.80\\ 
(without $c\bar{c}$) &&&&&&&&& \\ \hline
$  5.00-  5.50$~GeV
&  5.37&  5.41&  5.46&  5.40&  5.44&  5.50&  5.42&  5.47&  5.54\\ 
$  5.50-  6.00$~GeV
&  4.88&  4.91&  4.94&  4.89&  4.93&  4.97&  4.91&  4.95&  4.99\\ 
$  6.00-  9.46$~GeV
& 25.30& 25.37& 25.46& 25.37& 25.45& 25.55& 25.44& 25.53& 25.63\\ 
$  9.46- 10.52$~GeV
&  5.88&  5.89&  5.90&  5.90&  5.90&  5.91&  5.91&  5.92&  5.93\\ 
$  5.50- 10.52$~GeV
& 36.06& 36.17& 36.30& 36.16& 36.28& 36.43& 36.26& 36.39& 36.55\\ 
\hline\hline
$M_b$ (GeV) & $4.4$&$4.7$&$5.0$&$4.4$&$4.7$&$5.0$&$4.4$&$4.7$&$5.0$ \\ \hline
Energy range &&&&&&&&& \\ \hline
$ 10.52- 11.50$~GeV
&  4.94&  4.94&  4.95&  4.95&  4.95&  4.96&  4.96&  4.96&  4.97\\ 
(without $b\bar{b}$) &&&&&&&&& \\ \hline
$ 11.50- 12.00$~GeV
&   2.61&   2.61&   2.61&   2.62&   2.62&   2.62&   2.63&   2.63&   2.63\\ 
$ 12.00- 12.50$~GeV
&   2.51&   2.51&   2.51&   2.51&   2.51&   2.51&   2.52&   2.52&   2.52\\ 
$ 12.50- 13.00$~GeV
&   2.41&   2.41&   2.41&   2.42&   2.42&   2.42&   2.42&   2.42&   2.42\\ 
$ 13.00- 40.00$~GeV
&  72.78&  72.80&  72.81&  72.90&  72.92&  72.94&  73.03&  73.04&  73.06\\ 
$ 40.00-\infty$~GeV
&  42.61&  42.61&  42.62&  42.67&  42.67&  42.67&  42.73&  42.73&  42.73\\ 
$ 12.00-\infty$~GeV
& 120.31& 120.33& 120.34& 120.50& 120.52& 120.54& 120.70& 120.72& 120.74\\ 
\hline\hline
\end{tabular}
}
\parbox{14.cm}{\small
\caption{\label{tabdelallong}
Contributions to $\Delta\alpha^{(5)}_{\rm had}(M_Z^2) \times 10^4$ 
for different energy regions, quark masses and 
$\alpha_s^{(5)}(M_Z^2)$. The scale $\mu^2=s$ has been adopted.
Below the charm threshold the value $M_c=1.6$~GeV is chosen. 
In the energy ranges $3.73-5.00$~GeV and $10.52-11.50$~GeV for $R(s)$
the formulae valid below the corresponding quark threshold have been
used.
}}
\end{center}
\end{table}

\begin{table}[t]
\renewcommand{\arraystretch}{1.3}
\begin{center}
{\scriptsize
\begin{tabular}{|l||r|r|r|r|}
\hline\hline 
$\mu$ & $M_c$ & $2M_c$ & $\sqrt{s}$ & $2\sqrt{s}$\\ \hline\hline
Energy range &&&& \\ \hline
$   5.00-   5.50$~GeV
&   5.37&   5.44&   5.44&   5.43\\ 
$   5.50-   6.00$~GeV
&   4.85&   4.92&   4.93&   4.92\\ 
$   6.00-   9.46$~GeV
&  25.10&  25.35&  25.45&  25.46\\ 
$   9.46-  10.52$~GeV
&   5.85&   5.88&   5.90&   5.91\\ 
$   5.50-  10.52$~GeV
&  35.80&  36.14&  36.28&  36.28\\ 
\hline\hline
$\mu$ & $M_b$ & $2M_b$ & $\sqrt{s}$ & $2\sqrt{s}$\\ \hline\hline
Energy range &&&& \\ \hline
$  11.50-  12.00$~GeV
&   2.62&   2.62&   2.62&   2.62\\ 
$  12.00-  12.50$~GeV
&   2.52&   2.51&   2.51&   2.51\\ 
$  12.50-  13.00$~GeV
&   2.42&   2.42&   2.42&   2.41\\ 
$  13.00-  40.00$~GeV
&  72.89&  72.91&  72.92&  72.92\\ 
$  40.00-\infty$~GeV
&  42.63&  42.65&  42.67&  42.67\\ 
$  12.00-\infty$~GeV
& 120.45& 120.49& 120.52& 120.52\\ 
\hline\hline
\end{tabular}
}
\parbox{14.cm}{\small
\caption{\label{tabdelalmu}
Contribution to $\Delta\alpha^{(5)}_{\rm had}(M_Z^2) \times 10^4$
for different energy regions and different
choices of $\mu$ 
for the most heavy quark contribution (for the remaining
light quarks the scale $\mu^2 =s$ has been adopted).  
The values $\alpha_s^{(5)}(M_Z^2)=0.118$, $M_c=1.6$~GeV
and $M_b=4.7$~GeV have been chosen.
}}
\end{center}
\end{table}

\begin{table}[t]
\renewcommand{\arraystretch}{1.3}
\begin{center}
{\scriptsize
\begin{tabular}{|r|r|r|}
\hline
\hline
Energy range 
&
this work
&
Ref.~\cite{EidJeg95}
\\ \hline
$5.00-9.46$~GeV
&
$ (35.32 - 35.81)   \pm 0.4 $
&
$32.63$
\\
$12.00-40.00$~GeV
&
$ (77.82 - 77.84)   \pm  0.2 $
&
$79.22$
\\
\hline
\hline
\end{tabular}
}
\parbox{14.cm}{\small
\caption{\label{tabdelal}
Comparison of $\Delta\alpha^{(5)}_{\rm had}(M_Z^2) \times 10^4$
evaluated in this work with the data analyzed 
in~\protect\cite{EidJeg95}. The scale $\mu^2=s$ has been chosen
for the respective  light quark contributions and varied in the
range between $M_Q^2$ and $s$ for the most heavy quark. 
The errors arise from the variation of $M_c$, $M_b$ and
$\alpha_s^{(5)}(M_Z^2)$ (see Tab.~\ref{tabdelallong}).
}}
\end{center}
\end{table}

The detailed phenomenological analyses in 
Refs.~\cite{EidJeg95,MarZep95BurPie95Swa95}
have made use of the full set of data obtained by many different
experiments for energies from just above the two pion threshold up to
$40$~GeV.
Although different prescriptions  for the interpolation have been used
in the different papers, the most recent results are in fair
agreement. In the high energy region (typically
above $40$~GeV~\cite{EidJeg95}) the prediction 
for $R(s)$ based on perturbative QCD with
massless quarks has been employed.
A significant part of the final error originates from the region 
where perturbation theory should be reasonably valid: 
the light quark continuum, say above $2$~GeV,
the continuum above the charmonium resonances and below the
bottom threshold and the region between 12 and 40~GeV,
i.e. above the Upsilon resonances.
In view of the results presented in the
previous chapters one may employ perturbative QCD also in these regions.
This might lead to a reduction of the error,
albeit at the price of a more pronounced dependence on perturbative QCD.
(For early studies along this line see 
also~\cite{EidJeg95,MarZep95BurPie95Swa95}.) 

In Tab.~\ref{tabdelallong} and \ref{tabdelalmu}
the contributions to 
$\Delta\alpha^{(5)}_{\rm had}$
are displayed for a variety of input parameters
and renormalization scales. Since the validity of
our perturbative treatment is more doubtful
just above the respective charm and bottom thresholds, the contributions
from a variety of intervals are displayed separately. 
Once improved data are available, these theoretical
numbers could be replaced by more precise experimental
ones.
In Tab.~\ref{tabdelal} our results 
are compared to the analysis of~\cite{EidJeg95} for two energy ranges.
For the lower energy interval from $5.00 - 9.46$~GeV our 
QCD prediction for $\Delta\alpha^{(5)}_{\rm had}$ is bigger than
the value obtained by including experimental data by about
9\%. In contrast, for the energy interval from $12 - 40$~GeV, 
perturbative QCD gives a value slightly smaller than
the one obtained from integrating the experimental 
$R(s)$ values. This behaviour is already clear from 
Fig.~\ref{figrtot}, where we indicate the range of experimental data in
comparison to $R(s)$ from perturbative QCD.

If one subtracts the narrow
Upsilon resonances, QCD can be applied even up to the
threshold for $B$ meson production. In fact, recent experimental
investigations at 10.52~GeV~\cite{CLEO} are nicely consistent with
perturbative QCD. The additional contributions 
from the continuum cross section in the intervals between 
$3.73$~GeV and $5.00$~GeV (without $c\bar{c}$) and between
$9.46$~GeV and $10.52$~GeV (without $b\bar{b}$) are also listed 
in Tab.~\ref{tabdelallong}.
For the perturbative contributions through the range 
from $\sqrt{s}=3$~GeV to $\infty$
without charmonium and open charm contributions below $5$ GeV and
$\Upsilon$ and open bottom contributions below 11.5 GeV 
one finds
\begin{eqnarray}
\Delta\alpha^{(5)}_{\rm had}|_{\rm pert}
&=& 
\left[(186.27 - 186.87) \pm 0.70\right]\times 10^{-4}
\end{eqnarray}
where the range is due to the variation of $\mu$ for
the heavy quark contribution and the error due to 
uncertainties in the parameters. 
A more detailed discussion of the impact of these calculations will be 
given elsewhere.


\section{The region for small $\beta$ -- closer to the threshold}
\label{secthr}

In the high energy limit, say down to $s \approx 8M_Q^2$, the
cross section in ${\cal O}(\alpha_s)$ and ${\cal O}(\alpha_s^2)$ 
is well described by the massless approximation plus the
leading terms of the expansion in $M_Q^2/s$ up to 
$(M_Q^2/s)^6$~\cite{rhad,CheHarKueSte97}. The bulk of the
large logarithms is resummed by taking $\,\mu^2 = s\,$ for
the renormalization scale and adopting the $\overline{\rm MS}$
definition of the running mass. 
The fixed order result as given above is evidently adequate in the
intermediate energy region, with the requirement that
$C_F\pi\alpha_s/\beta$ is not yet too large, i.e. safely away from
the threshold regime, where the conventional multi-loop expansion
breaks down.
Otherwise some care
has to be taken to control the higher order terms proportional to
$(C_F\pi\alpha/\beta)^n$ with $n\ge 0$. As far as dominant and
subdominant contributions of this sort are concerned, their structure
is understood from nonrelativistic considerations and will be briefly
outlined in the following for the case $C_F\pi\alpha_s \lsim \beta\ll1$.

Let us in a first step discuss those terms which are multiplied by the
colour factor $C_F$ only and which are relevant for QCD and QED. For
the dominant contributions to the cross section in the nonrelativistic
limit, often called ``Sommerfeld factor'' in literature, the leading
terms in an expansion in $x_S=C_F\pi\alpha_s/\beta$ are given by
\begin{eqnarray}
R^{\rm thr} &=& 3Q_Q^2\frac{3}{2}\frac{\beta x_S}{1-e^{-x_S}} 
\nonumber\\
&=& 
3Q_Q^2\frac{3}{2}\,\beta\left(1+\frac{x_S}{2}
                          +\frac{B_1 x_S^2}{2!}
                          -\frac{B_2 x_S^4}{4!}
                          +\frac{B_3 x_S^6}{6!}
                          -\ldots
                          +(-1)^{(n+1)}\frac{B_n x_S^{2n}}{(2n)!}
                          \pm\ldots
                   \right)
\nonumber\\
    &=& 3Q_Q^2\frac{3}{2}\,\beta\left(
       1 + C_F\frac{\alpha_s}{\pi} \frac{\pi^2}{2\beta}
         + C_F^2\left(\frac{\alpha_s}{\pi} \right)^2 \frac{\pi^4}{12 \beta^2}
         + \ldots
       \right),
\label{eqthresh1}
\end{eqnarray}
where $B_n$ are the Bernoulli numbers: 
$B_1=1/6,\, B_2=1/30,\, B_3=1/42,\,\ldots\,\,\,$.
It should be noted that the Sommerfeld factor is entirely of
long-distance origin and proportional to the imaginary part of the
nonrelativistic Green function for the Coulomb
potential~\cite{Bra68}, i.e. governed by the continuum Coulomb wave function.
These terms are predicted from a consideration, where the $Q\bar Q$ 
production process is decomposed into a short distance part (to be 
eventually corrected by short distance QCD corrections) and a long 
distance part, which is governed by the Coulomb wave function, in other 
words, by the imaginary part of the nonrelativistic Green function for
the Coulomb potential~\cite{Vol79,Leu81}.

Resummed and fixed order results have to coincide in the region of
small $x_S$. Thus it is instructive to compare the
Sommerfeld factor and the sum of the first three terms in the
expansion~(\ref{eqthresh1}) 
(corresponding to Born, one- and two-loop contributions,
respectively) for various values of $x_S$ which are not too much larger than
one. It is remarkable that the sum of the first
three terms in Eq.~(\ref{eqthresh1}) provides an excellent approximation not 
only for small values of $x_S$, but even up to $x_S = 2$ with a
relative deviation of less than 1\%. Even for $x_S = \pi$, corresponding 
to $C_F\alpha_s/\beta = 1$, the deviation amounts to 3.3\% only. 

In addition, based on the consideration that the cross section for
nonrelativistic energies can be decomposed into long- and
short-distance contributions
one obtains in QED an additional 
correction factor which comes from relativistic momenta involved in
the transverse photon exchange. This factor, which is quite familiar
from the single-photon annihilation contributions to the positronium
hyperfine splitting~\cite{KarKle52} and from the corrections to
quarkonium annihilation through a virtual photon~\cite{BarGatKoeKun75}, 
can be derived from the threshold behaviour of the one-loop 
corrections~\cite{KaeSab55}. In
combination one thus anticipates the following behaviour
\begin{equation}
R^{\rm thr} = 3Q_Q^2\beta\,\frac{3-\beta^2}{2} \Bigg[ 1 + 
 \frac{1}{2}\,C_F\frac{\alpha_s\pi}{\beta}\,\left(1 + \beta^2\right) + 
 \frac{1}{12}\left(C_F\frac{\alpha_s\pi}{\beta}\,
                   \left(1 + \beta^2\right)\right)^2 \Bigg] 
 \left(1 - 4C_F\,\frac{\alpha_s}{\pi}\right)\,.
\label{eqthresh2}
\end{equation}
The inclusion of the
subleading $\beta^2$ terms in the combinations $(3-\beta^2)$ and
$(1+\beta^2)/\beta$ is suggested from the structure of Eq.~(\ref{eqr1}) 
for small $\beta$. 
An explicit proof for this structure at the NLO
level can be found in~\cite{Hoa972}.
Their interpretation as long distance contributions is furthermore
motivated by the appearance of the logarithmic terms with the same
structure in the $C_F T n_\ell$ contributions listed below 
(see~\cite{CHKST1} and Eq.~(40) of~\cite{CKS1}).
Even the impact of the running of the coupling
constant can be included in this nonrelativistic line of reasoning. In
QCD the running of $\alpha_s$ is induced by terms proportional to
$T\,n_{\ell}$ and $C_A$. 

For a prediction of the hadronic cross section very close to
threshold, i.e. in the regime $\beta\le C_F\pi\alpha_s$, the
factorization into short- and long-distance contributions analogous to
Eq.~(\ref{eqthresh2}) is desirable, incorporating also the 
new information from the NLO
calculation. Such an analysis requires the resummation of
long-distance contributions to all orders and shall not be carried out
here\footnote{
A resummation of this sort for the QED
contributions, i.e. including also the order $\beta$ terms in 
Eq.~(\ref{eqthresh2}), can be found in~\cite{Hoa97}.
}.
However, the leading terms of the perturbative series for
small $\beta$ provide the basis for such a resummation and will be
collected in the following.

Let us now recapitulate the threshold behaviour of the various 
ingredients for $R_c$:
\begin{eqnarray}
R_c^{(0)} & = & 3Q_c^2\,\beta \frac{3-\beta^2}{2},
\\
R_c^{(1)} & = & R^{(0)}  
\left[\frac{\pi^2(1+\beta^2)}{2\beta}-4\right] 
+ \ldots\,\,,\\
R_A^{(2)} & = & R^{(0)}  
\left[\frac{1}{12}\left(\frac{\pi^2(1+\beta^2)}{\beta}\right)^2
-2\frac{\pi^2(1+\beta^2) }{\beta}
\right.\nonumber\\&&\left.\quad\mbox{}
 -\frac{2}{3} \pi^2\left(\ln\frac{\beta}{4}+\frac{35}{12}\right) 
            +\frac{39}{4}-\zeta(3)   
\right] 
+ \ldots\,\,,
\label{eqthrra}\\
R_{NA}^{(2)} & = & R^{(0)} 
\left[\frac{\pi^2(1+\beta^2)}{\beta}
\left(-\frac{11}{24}\ln\frac{4\beta^2 M_c^2}{\mu^2}+\frac{31}{72}\right)
-\frac{11}{3}\ln\frac{\mu^2}{M_c^2} + c
\right]
+ \ldots\,\,,
\label{eqthrrna}\\
R_{cq}^{(2)} & = & R^{(0)} 
\left[\frac{\pi^2(1+\beta^2)}{\beta}
\left(\frac{1}{6}\ln\frac{4\beta^2 M_c^2}{\mu^2}-\frac{5}{18}\right)
+\frac{4}{3}\ln\frac{\mu^2}{M_c^2}+\frac{11}{9}
\right] 
+ \ldots\,\,,
\label{eqrcell}\\
R_{cc}^{(2)} & = & R^{(0)} 
\left[
\frac{\pi^2}{6\beta} \ln\frac{M_c^2}{\mu^2} 
-\frac{4}{3}\ln\frac{M_c^2}{\mu^2} 
+\frac{44}{9}-\frac{8}{3}\zeta(2)
\right]
+\ldots
\,\,.
\label{eqrthresholdparts}
\end{eqnarray}

Terms of order $\beta^2$ (modulo $\ln\beta$) are neglected.
The order $\beta$ terms in Eq.~(\ref{eqthrra}) 
is taken from~\cite{Hoa97}. 
Subleading terms in Eq.~(\ref{eqthrrna}), symbolized by $c$, have
not been calculated yet analytically. Assuming a linear dependence on
$\beta$, an estimate for the constant $c$ can be extracted from the
numerical analysis in~\cite{CKS1}.
Considering the expansion of $11$ different Pad\'e
approximations which show a quite stable behaviour near threshold
one obtains
\begin{eqnarray}
c &=& 24 \pm 5
\,,
\end{eqnarray}
where the error is estimated by taking into account the variation of
the predictions for the constant $c$ from the different Pad\'e
approximations. 
This numerical result for $c$ is fairly large,
in particular when compared to the corresponding constant $11/9$
in the $C_FTn_\ell$ term, Eq.~(\ref{eqrcell}).

In total one thus gets\footnote{
The  contributions  from  $\delta R^{(2)}_{cb}$ are
suppressed by the power $M_c^2/M_b^2$, not Coulomb enhanced
and thus ignored in the following.
}:
\begin{eqnarray}
R_c &=& Q_c^2\Bigg[
R_c^{(0)}
+\frac{\alpha_s^{(4)}(\mu^2)}{\pi} C_FR^{(1)}
\nonumber\\
&&+\left(\frac{\alpha_s^{(4)}(\mu^2)}{\pi}\right)^2
\left(
C_F^2 R_A^{(2)} + C_F C_A R_{NA}^{(2)} +
T C_F n_{\ell} R_{c q}^{(2)} + T C_F R_{c c}^{(2)}
\right)
\Bigg].
\label{eqrc012a}
\end{eqnarray}
The renormalization group invariance of this result is apparent:
the $\mu$ dependence of $R_{NA}$, $R_{cq}$,  and $R_{cc}$
properly compensates the $\mu$ dependence of the ${\cal O}(\alpha_s)$
term.
As stated above, the energies considered in this section will be of order 
$M_c$, implying that $\ln(s/M_c^2)$ is not a large quantity. The transition 
from $\alpha_s^{(n_f)}$ to $\alpha_s^{(n_{\ell})}$ is thus legitimate and 
easily achieved by absorbing the last term of Eq.~(\ref{eqr2cc}) in 
the order $\alpha_s$ expression. 

The logarithmic singularities and the constants 
in $R_{NA}$ and $R_{cq}$ which are  leading in $\beta$ 
can be absorbed in the
terms of order $\alpha_s$ if the $\overline{\rm MS}$ coupling constant
is replaced by the coupling governing the potential~\cite{Fis77}
\begin{eqnarray}
 V_{QCD}(\vec{q}\,^2) &=& 
         -4\pi C_F\frac{\alpha_V(\vec{q}\,^2)}{\vec{q}\,^2},
\\
 \alpha_V(\vec{q}\,^2) &=&  \alpha_s^{(n_\ell)}(\mu^2)\Bigg[
      1 + \frac{\alpha_s^{(n_\ell)}(\mu^2)}{4\pi}\left(
          \left(\frac{11}{3}C_A-\frac{4}{3}T n_\ell\right)
          \left(-\ln\frac{\vec{q}\,^2}{\mu^2}+\frac{5}{3}\right)
          -\frac{8}{3}C_A            \right)         
\nonumber
\Bigg],
\end{eqnarray}
with $\vec{q}\,^2 = \beta^2 s$. This is apparent once the sum of the Born 
cross section plus higher order corrections is rewritten as follows
(with $x_V = C_F\pi\alpha_V(\beta^2s)\,(1+\beta^2)/\beta$):
\begin{eqnarray}
R_c &=& Q_c^2\,R_c^{(0)}\Bigg\{
1+\frac{x_V}{2}
-4C_F\frac{\alpha_s^{(3)}(\mu^2)}{\pi}
+\frac{x_V^2}{12}
-4C_F\frac{x_V}{2}\frac{\alpha_s^{(3)}(\mu^2)}{\pi}
\nonumber\\&&
+\left(\frac{\alpha_s^{(3)}(\mu^2)}{\pi}\right)^2
\Bigg[
C_F^2\left( 
 -\frac{2}{3} \pi^2\left(\ln\frac{\beta}{4}+\frac{35}{12}\right) 
            +\frac{39}{4}-\zeta(3)   
\right)
\nonumber\\&&\mbox{}
+C_AC_F\left(
-\frac{11}{3}\ln\frac{\mu^2}{M_c^2}+ c
\right)
+C_FTn_\ell\left(
\frac{4}{3}\ln\frac{\mu^2}{M_c^2}+\frac{11}{9}
\right)
\nonumber\\&&\mbox{}
+C_FT\left(
\frac{44}{9}-\frac{8}{3}\zeta(2)
\right)
\Bigg]
\Bigg\}
+ \ldots\,\,.
\label{eqrc012b}
\end{eqnarray}
In the transition from Eq.~(\ref{eqrc012a}) to 
(\ref{eqrc012b}) we have freely dropped terms of
order $\alpha_s^3$. It is evident that the scale in the correction term
from hard transverse gluon exchange
proportional to $4 C_F \alpha_s^{(3)}(\mu^2) / \pi$
is of order $M_c$, with $\mu_{BLM} = e^{-11/24} M_c = 0.63 M_c$
suggested by the BLM prescription~\cite{BroLepMac83}. 
However, as noted before, the corresponding constant
$c\approx24$ in the non-abelian term is markedly different from
what would be expected from the BLM procedure.

As stated above these results
are strictly applicable in the limit $\pi C_F \alpha_s 
\lsim \beta \ll 1$ only. 
Nevertheless, Eq.~(\ref{eqrc012b}) provides an important input 
for the determination
of the cross section very close to threshold and even for bound state
energies, i.e. for $|\beta| < C_F\pi\alpha_s$, because it contains all
${\cal O}(\alpha_s^2)$ short-distance effects relevant for the
nonrelativistic regime. These short-distance effects, which are
specific for the single-photon annihilation process involving massive
quark-antiquark pairs, are universal for $|\beta|\ll 1$ regardless
whether $|\beta|$ is smaller or larger than $C_F\pi\alpha_s$. In this
respect Eq.~(\ref{eqrc012b}) even provides an important result for the
investigation of leptonic decay widths of the $\Psi$ and the
$\Upsilon$ families, and for QCD sum rules for the $b\bar b$ system.
A discussion of the latter subjects, however, is beyond the scope of
this paper.


\section{Summary and Conclusions}
\label{secsum}

Predictions for the cross section of massive quark production in
$e^+e^-$ annihilation are presented which are accurate to order
$\alpha_s^2$. Their range of validity extends from high energies down
close to threshold, i.e. to center of mass 
energies of about 5~GeV, 11.5~GeV and
$2 M_t + 12$~GeV for charm, bottom and top respectively. Inclusion of
the leading and subleading terms of order $\alpha_s^3$
proportional to $M_Q^2/s$ 
allows to connect smoothly the NNLO prediction at high energies with the
NLO prediction in the intermediate and ``low'' energy range. The NLO
corrections are sizable and must be taken into account to achieve a
prediction with an accuracy better than 10\%. The stability of the
prediction against variations of the renormalization scale and of the
input values for the quark masses and $\alpha_s$ has been tested.
Even fairly extreme assumptions about the renormalization scale lead to
moderate variations in the case of charm and to negligible variations
for the heavier quarks. The same holds true for the dependence on the
input parameters, the quark masses and the strong coupling constant. 
Only a change in the charm quark mass has a clearly
visible effect on the cross section. Within the large experimental
errors, theoretical and experimental results are well consistent. The
theoretical results can now be used to perform an universal
fit to the data, including the lower energy regions. In view of the
sparse data with large errors in the region from $2 - 3.73$~GeV,
from $5 - 10.52$~GeV
and from $11.5 - 40$~GeV one
could also consider to use the predictions based on
perturbative QCD to
arrive at a more precise value for the running QED coupling at the $Z$
boson mass. A significant reduction of the uncertainty could be
obtained.

\vspace{10mm}
\noindent
{\large\bf Acknowledgments}

\medskip

\noindent
We thank John Outhwaite for providing us with his
compilation of the experimental data for the inclusive $R$
measurements. T.~Teubner thanks the British PPARC and the University 
of Durham, where part of this work was carried out.
K.G.~Chetyrkin appreciates
the warm hospitality of the Theoretical group of the Max Planck
Institute in Munich where part of this work has been made.
A.H.~Hoang is supported in part by U.S. Department of Energy under Contract 
No. DOE DE-FG03-90ER40546.

\end{document}